# Modularity Aspects of Disjunctive Stable Models

**Tomi Janhunen**                                    Tomi.Janhunen@tkk.fi
**Emilia Oikarinen**                                 Emilia.Oikarinen@tkk.fi
*Helsinki University of Technology*
*Department of Information and Computer Science*
*P.O. Box 5400, FI-02015 TKK, Finland*

**Hans Tompits**                                     tompits@kr.tuwien.ac.at
**Stefan Woltran**                                   woltran@dbai.tuwien.ac.at
*Technische Universität Wien*
*Institut für Informationssysteme*
*Favoritenstraße 9–11, A-1040 Vienna, Austria*

## Abstract

Practically all programming languages allow the programmer to split a program into several modules which brings along several advantages in software development. In this paper, we are interested in the area of answer-set programming where fully declarative and nonmonotonic languages are applied. In this context, obtaining a modular structure for programs is by no means straightforward since the output of an entire program cannot in general be composed from the output of its components. To better understand the effects of disjunctive information on modularity we restrict the scope of analysis to the case of disjunctive logic programs (DLPs) subject to stable-model semantics. We define the notion of a *DLP-function*, where a well-defined input/output interface is provided, and establish a novel module theorem which indicates the compositionality of stable-model semantics for DLP-functions. The module theorem extends the well-known splitting-set theorem and enables the decomposition of DLP-functions given their strongly connected components based on positive dependencies induced by rules. In this setting, it is also possible to split shared disjunctive rules among components using a generalized shifting technique. The concept of modular equivalence is introduced for the mutual comparison of DLP-functions using a generalization of a translation-based verification method.

## 1. Introduction

Practically all programming languages used in software development allow the programmer to split a program into several modules which interact through well-defined input/output interfaces. Given this, the entire program can be viewed as a *composition* of its component modules which are typically linked together in the respective run-time environment. The expected benefits of modular program development are manifold. First, it imposes a good programming style to be followed by the programmer. A complex software system is much easier to develop as a set of interacting components rather than a monolithic program. Second, a modular architecture allows for additional flexibility as regards delegating programming tasks amongst a team of programmers. In this setting, the goal of each programmer is to implement desired input/output behavior(s) in terms of concrete module(s) which together implement the software system being developed. Third, modular program design can also be exploited in order to boost the execution of programs. Program optimization is also facilitated by structural information encompassed by module interfaces.

*Answer-set programming* (ASP) (Marek & Truszczyński, 1999; Niemelä, 1999; Gelfond & Leone, 2002) is a paradigm for declarative problem solving in which solutions of problems are described in terms of *rules* subject to a nonmonotonic semantics based on stable models (Gelfond & Lifschitz, 1988). In typical problem representations, a tight correspondence between solutions and stable models is sought for, and *default negation* is fully exploited in order to obtain concise encodings of relations involved in such problem descriptions. Furthermore, recursive definitions enable, e.g., the representation of *closures* of relations in a very natural way. Due to efficient implementations and emerging applications, the paradigm has received increasing attention during the past two decades.[1] In the meantime, a number of extensions—such as *disjunctions*, *weight constraints*, and *aggregates*—have been proposed to the basic syntax of *normal logic programs*. In this paper, we concentrate on the class of *disjunctive logic programs* (DLPs) which is appropriate for solving search problems residing up to the second level of the polynomial-time hierarchy. The semantical account of DLPs is based on the respective generalization of stable-model semantics (Gelfond & Lifschitz, 1991).

In this paper, our goal is to investigate modularity in the context of DLPs and stable-model semantics. Since stable models are defined only over complete programs, they do not lend themselves to modular programming *prima facie*. Perhaps for this reason, the concept of a module has not yet raised too much attention in the realm of answer-set programming. Except for a few dedicated papers (Gaifman & Shapiro, 1989; Eiter, Gottlob, & Veith, 1997b; Baral, Dzifcak, & Takahashi, 2006), modules mostly appeared as a by-product in studies of formal properties like stratification, splitting, or, more lately, in work on equivalence relations between programs (Lifschitz & Turner, 1994; Eiter, Gottlob, & Mannila, 1997a; Eiter, Ianni, Lukasiewicz, Schindlauer, & Tompits, 2008). In a recent approach of Oikarinen and Janhunen (2008a), the modular architecture put forth by Gaifman and Shapiro (1989) is accommodated for the classes of normal and SMODELS programs. The main result is a *module theorem* which links stable models associated with individual modules to those of their *composition*. Such a result is significant as it indicates that stable models are compositional in very much the same sense as classical models are in propositional logic. The only major restriction implied by the module theorem is that the definition of any set of *positively interdependent* atoms must be given within the same module.

Besides the general benefits of modular program development discussed above, we are also looking for potential computational advantages of modularizing reasoning tasks in ASP. In this context, the search for stable models is probably the most central reasoning task. Results like the module theorem discussed above provide the basis for modularizing the search task. Extra care, however, is required because the computation of stable models for modules in separation is not necessarily efficient. More sophisticated methods, such as identifying *cones of influence* in Boolean circuits (Junttila & Niemelä, 2000), can be devised to identify modules which are relevant for the search of stable models—the rest is only used to expand a qualified stable model to one for the entire program. This strategy alleviates the treatment of extremely large program instances and it is also amenable to query evaluation.

---

1. The 20th anniversary of stable-model semantics was celebrated at ICLP'08 which was held in Udine, Italy, in December 2008.

Unfortunately, contemporary disjunctive answer-set solvers, such as CLASPD (Drescher et al., 2008), CMODELS (Giunchiglia, Lierler, & Maratea, 2006), DLV (Leone et al., 2006), and GNT (Janhunen, Niemelä, Seipel, Simons, & You, 2006), exhibit little support for modular reasoning although related techniques like strongly connected components are exploited internally. There are also other reasoning tasks that can be boosted with a modular approach. For instance, the optimization of answer-set programs gives rise to the problem of *verifying* whether different versions of programs have the same answer sets. As demonstrated by Oikarinen and Janhunen (2009), such verification tasks may benefit from modularization, and, in particular, if approximation techniques based on *modular equivalence* are introduced. Following this idea, the first modular off-line optimizer of answer-set programs, called MOD-OPT, has recently been implemented (Janhunen, 2008b).

There are also other interesting applications of modules in sight: Gebser et al. (2008a) propose an *incremental* technique for answer-set solving. The idea is to gradually extend a program instance in terms of additional modules, e.g., when solving AI planning problems. Moreover, theoretical results like the splitting-set theorem (Lifschitz & Turner, 1994) and the module theorem can be directly exploited in correctness proofs. For instance, it is proved by Oikarinen and Janhunen (2008b) that the models of a prioritized circumscription can be captured with disjunctive stable models using a particular translation. A similar proof strategy is adopted in Theorem 8.5 of this paper.

We anticipate that compositional semantics can also prove useful if one tries to boost the search for stable models via parallelization, e.g., by computing stable models for modules in parallel. However, in order to avoid excessive communication costs, extra caution is needed when stable models computed in separation are linked together and potentially rejected. One possibility is to identify mutually independent modules as the basis for distribution. Besides this aspect, modularization may also lead to novel methods for the (non-parallelized) computation of stable models, other than the traditional ones.

**Structure and Preview of Results** In this paper, we concentrate on the formal underpinnings of modular programming in the context of disjunctive logic programs under stable-model semantics. We proceed as follows. Our first goal is to generalize the theory developed for normal programs and SMODELS programs (Oikarinen & Janhunen, 2008a) to the case of disjunctive programs. To this end, we first introduce the notion of a *DLP-function* in Section 2. The term goes back to Gelfond and Gabaldon (1999) who introduced *LP-functions* as (partial) definitions of new relations in terms of old, known ones. To enable such a functional view of disjunctive programs, they are endowed with a well-defined input/output interface. The idea is to partition the signature of a program encapsulated in this way into *input atoms*, *output atoms*, and *hidden* (or *local*) *atoms*. These distinctions provide the basis for the systematic *composition* of larger disjunctive logic programs out of program modules. However, arbitrary combinations of program modules are not meaningful and, first of all, we adopt syntactic restrictions introduced by Gaifman and Shapiro (1989) from the context of negation/disjunction-free logic programs. The interplay of default negation and disjunctions brings along new factors which lead to a relaxation of the restrictions in the sense that program modules are allowed to *share* rules. Then, having the basic syntactic issues of DLP-functions laid out, we concentrate on their semantics in Section 3. In this respect, we follow a strict model-theoretic approach and, in particular, address the role of input atoms when it comes to viewing DLP-functions as mathematical functions. We proceed step by step and assign three different classes of models to each DLP-function, viz. classical models, minimal models, and stable models. The last provides an appropriate generalization of disjunctive stable models (Gelfond & Lifschitz, 1991) in the presence of input atoms.

Our second objective is to establish the adequacy of the concept of a DLP-function in view of a *compositional* semantics. This will be witnessed by the main result of the paper, viz. the *module theorem* which shows how stable models of a DLP-function, $\Pi$, can be alternatively obtained as unions of *compatible* stable models for the modules constituting $\Pi$. The proof of the theorem is based on the notions of *completion* (Clark, 1978) and *loop formulas* (Lin & Zhao, 2004; Lee & Lifschitz, 2003) which are first lifted to the case of DLP-functions in Section 4 as a preparatory step. The proof of the module theorem follows as the main topic of Section 5. The result is non-trivial because the underlying semantics based on stable models is inherently nonmonotonic. This feature was already recognized by Gaifman and Shapiro (1989) in a much simpler setting of definite programs—neither involving default negation nor disjunctions. As observed by them, too, syntactic restrictions on program composition are necessary in order to guarantee compositionality properties for the semantics based on Herbrand models.[2] In the current paper, we strive for analogous results but in the case of programs permitting both default negation and disjunctions. It turns out that *strongly connected components* of positive dependency graphs provide a key criterion when it comes to confining program composition. The compositionality properties of disjunctive programs under stable-model semantics have also arisen in the context of the so-called *splitting-set theorem* (Lifschitz & Turner, 1994; Eiter et al., 1997a, 2008). In fact, the module theorem established herein is a proper generalization of its predecessor (Oikarinen & Janhunen, 2008a). We illustrate the potential of our modular architecture by the evaluation of quantified Boolean formulas (QBFs), which serve as canonical representatives of the classes of the polynomial-time hierarchy (PH). Due to basic complexity results established by Eiter and Gottlob (1995), it is natural from our perspective to concentrate on the second level of the PH in the case of disjunctive programs.

The third aim of this paper is to have a look at some particular applications of the module theorem in disjunctive logic programming. In Section 6, we take an opposite view to the modular construction of DLP-functions and consider possibilities for their *decomposition* even in the absence of any other structural information. It turns out that strongly connected components can also be exploited in this respect but, in addition, the occurrences of hidden atoms must be taken into account when splitting a DLP-function into its components. As demonstrated in Section 7, our results open new prospects as regards unwinding disjunctions using the principle of *shifting* (Gelfond, Przymusinska, Lifschitz, & Truszczyński, 1991; Dix, Gottlob, & Marek, 1996; Eiter, Fink, Tompits, & Woltran, 2004). A proper generalization of this principle that partially covers also programs involving *head-cycles* is formulated and proved correct. Moreover, due to the modular nature of DLP-functions, it makes perfect sense to compare them as modules. The notion of *modular equivalence* is introduced for this purpose in Section 8. Interestingly, modular equivalence supports substitutions of equivalent programs and it also lends itself for *translation-based verification* as put forth by Oikarinen and Janhunen (2004, 2009) in the related cases of *ordinary equivalence* and SMODELS *programs*. Section 9 contrasts our approach with related work. Finally, Section 10 provides a brief summary of results and concludes this paper.

---

2. The main concern of Gaifman and Shapiro (1989) is modularity with respect to the logical consequences of a definite program and hence the intersection of its Herbrand models.

## 2. The Class of DLP-Functions

The topic of this section is the syntax of DLP-functions as well as syntactic restrictions imposed on composition of DLP-functions. A *disjunctive rule* is an expression of the form

$$a_1 \vee \cdots \vee a_n \leftarrow b_1, \ldots, b_m, \sim c_1, \ldots, \sim c_k, \tag{1}$$

where $n, m, k \geq 0$, and $a_1, \ldots, a_n$, $b_1, \ldots, b_m$, and $c_1, \ldots, c_k$ are propositional atoms. Since the order of atoms is considered insignificant, we write $A \leftarrow B, \sim C$ as a shorthand for rules of form (1), where $A = \{a_1, \ldots, a_n\}$, $B = \{b_1, \ldots, b_m\}$, and $C = \{c_1, \ldots, c_k\}$ are the respective sets of atoms. The basic intuition behind a rule $A \leftarrow B, \sim C$ is that if each atom in the positive body $B$ can be inferred and none of the atoms in the negative body $C$, then some atom in the head $A$ can be inferred. When both $B$ and $C$ are empty, we have a *disjunctive fact*, written $A \leftarrow$. If $A$ is empty, then we have a *constraint*, written $\bot \leftarrow B, \sim C$.

A *disjunctive logic program* (DLP) is conventionally formed as a finite set of disjunctive rules. Additionally, we want a distinguished *input and output interface* for each DLP. To this end, we extend a definition originally proposed by Gaifman and Shapiro (1989) to the case of disjunctive programs.[3] It is natural that such an interface imposes certain restrictions on the rules allowed in a module. Given a set $R$ of disjunctive rules, we write $\text{At}(R)$ for the *signature* of $R$, i.e., the set of (ground) atoms effectively appearing in the rules of $R$.

**Definition 2.1** *A DLP-function,* $\Pi$*, is a quadruple* $\langle R, I, O, H \rangle$*, where $I$, $O$, and $H$ are pairwise distinct sets of* input atoms, output atoms, *and* hidden atoms, *respectively, and $R$ is a DLP such that for each rule $A \leftarrow B, \sim C \in R$,*

*1. $A \cup B \cup C \subseteq I \cup O \cup H$, and*

*2. if $A \neq \emptyset$, then $A \cap (O \cup H) \neq \emptyset$.*

A DLP-function $\Pi = \langle R, I, O, H \rangle$ is occasionally identified with $R$ and, by a slight abuse of notation, we write $A \leftarrow B, \sim C \in \Pi$ to denote $A \leftarrow B, \sim C \in R$. By the first condition of Definition 2.1, the rules in a DLP-function $\Pi$ must obey the interface specification of $\Pi$, i.e., $\text{At}(R) \subseteq I \cup O \cup H$. As regards the sets of atoms $I$, $O$, and $H$ involved in the module interface, the atoms in $I \cup O$ are considered to be *visible* and hence accessible to other DLP-functions conjoined with $\Pi$; either to produce input for $\Pi$ or to utilize the output of $\Pi$. On the other hand, the *hidden* atoms in $H$ are used to formalize some auxiliary concepts of $\Pi$ which may not make sense in the context of other DLP-functions but may save space substantially as demonstrated, e.g., by Janhunen and Oikarinen (2007, Example 4.5). The second condition of Definition 2.1 is concerned with the set of atoms $O \cup H$ *defined* by the rules of $R$. The principle is that each non-empty disjunctive head must involve at least one atom from $O \cup H$. This is just to ensure that a DLP-function $\Pi$ must not interfere with the definitions of its input atoms $I$ in terms of rules $A \leftarrow B, {\sim}C$ satisfying $A \subseteq I$. But otherwise, the rules of $\Pi$ may be conditioned by input atoms.[4] Given a set $S$ of atoms, we distinguish the set of rules that define the atoms of $S$ in $R$, i.e., the set of *defining rules*

$$\mathrm{Def}_R(S) = \{A \leftarrow B, {\sim}C \in R \mid A \cap S \neq \emptyset\}. \tag{2}$$

---

3. Similar approaches within the area of ASP have previously been introduced by Gelfond and Gabaldon (1999), Janhunen (2006), and Oikarinen and Janhunen (2008a).

Our next objective is to specify the conditions on which the composition of DLP-functions may take place. Roughly speaking, the idea is that larger DLP-functions can be formed in a modular fashion using smaller DLP-functions as components. As observed already by Gaifman and Shapiro (1989), syntactic restrictions on program composition are necessary in order to guarantee compositionality properties for the semantics based on Herbrand models, even for the simple case of definite programs. Thus, program union as operator for composition without further restrictions is not satisfactory with respect to compositionality.

We start by adapting the construction of Gaifman and Shapiro (1989) to the case of disjunctive programs.

**Definition 2.2** *Two DLP-functions* $\Pi_1 = \langle R_1, I_1, O_1, H_1 \rangle$ *and* $\Pi_2 = \langle R_2, I_2, O_2, H_2 \rangle$ *respect the input/output interfaces of each other if and only if*

1. $(I_1 \cup O_1 \cup H_1) \cap H_2 = \emptyset$,

2. $(I_2 \cup O_2 \cup H_2) \cap H_1 = \emptyset$,

3. $O_1 \cap O_2 = \emptyset$,

4. $\mathrm{Def}_{R_1}(O_1) = \mathrm{Def}_{R_1 \cup R_2}(O_1)$, *and*

5. $\mathrm{Def}_{R_2}(O_2) = \mathrm{Def}_{R_1 \cup R_2}(O_2)$.

The first three of the conditions above are due to Gaifman and Shapiro (1989) and they imply that the sets $O_1$, $H_1$, $O_2$, and $H_1$ are mutually pairwise distinct. Violations with respect to the first two conditions can be circumvented by a renaming strategy. For instance, if an atom $a \in H_1$ appears in $I_2 \cup O_2 \cup H_2$, hence violating the second condition, it is possible to replace all occurrences of $a$ in $\Pi_1$ by a new atom $a' \notin I_2 \cup O_2 \cup H_2$ not appearing in $\Pi_2$. This removes the conflict with respect to $a$ and so forth.[5]

On the other hand, the last two conditions of Definition 2.2 concern the distribution of rules involved in the *definitions* (2) of sets of atoms $O_1$ and $O_2$, i.e., the sets of rules $\mathrm{Def}_{R_1}(O_1)$ and $\mathrm{Def}_{R_2}(O_2)$, in $R_1$ and $R_2$, respectively. As regards disjunctive rules, the principle is that these sets of defining rules must remain intact when the union $R_1 \cup R_2$ is formed which means that each module is supposed to have copies of all rules that form the definition of its output atoms. In spite of this, two modules $\Pi_1$ and $\Pi_2$ subject to the conditions of Definition 2.2 may effectively *share* disjunctive rules $A \leftarrow B, {\sim}C$ with a non-empty head $A$ such that $A \cap O_1 \neq \emptyset$ and $A \cap O_2 \neq \emptyset$, as to be demonstrated next.

---

4. In particular, input atoms in the head $A$ of a rule act very much like atoms in the negative body ${\sim}C$.

5. An opposite view to program composition is considered in Section 6, where possibilities for decomposing a disjunctive program into smaller DLP-functions are studied. As a counterpart to renaming, a revealing operator introduced in Definition 7.3 can be used for circumventing the first two conditions in Definition 2.2.

**Example 2.3** *Consider the following two DLP-functions:*[6]

| {b} |
|---|
| $a \vee b \leftarrow c;$ |
| $d \leftarrow a, \sim d$ |
| {a, c} |

*and*

| {a} |
|---|
| $a \vee b \leftarrow c;$ |
| $e \leftarrow a, \sim e$ |
| {b, c} |

*More formally, we have $\Pi_1 = \langle R_1, \{a, c\}, \{b\}, \{d\}\rangle$ and $\Pi_2 = \langle R_2, \{b, c\}, \{a\}, \{e\}\rangle$ such that $R_1 \cap R_2 = \{a \vee b \leftarrow c\}$. We show that $\Pi_1$ and $\Pi_2$ respect the input/output interfaces of each other: First, both hidden atoms $d$ and $e$ occur in exactly one of the two programs and thus the first two conditions in Definition 2.2 are satisfied. Second, we have disjoint output atoms, viz. atom $b$ in $\Pi_1$ and atom $a$ in $\Pi_2$. Finally, we have $\mathrm{Def}_{R_1}(\{b\}) = \mathrm{Def}_{R_1 \cup R_2}(\{b\}) = \mathrm{Def}_{R_2}(\{a\}) = \mathrm{Def}_{R_1 \cup R_2}(\{a\}) = \{a \vee b \leftarrow c\}$, which shows that also the final two conditions in Definition 2.2 are satisfied, and as far as syntax is concerned, it makes sense to compose a larger DLP-function which is obtained as a kind of a union of $\Pi_1$ and $\Pi_2$; see (4) below.*

In contrast to disjunctive programs, shared rules do not arise in the context of normal logic programs since only one head atom is allowed in each rule. The same can be stated about SMODELS programs (Simons, Niemelä, & Soininen, 2002) although such programs may contain, among other rule types, *choice rules* of the form

$$\{a_1, \ldots, a_n\} \leftarrow B, \sim C \tag{3}$$

with heads of cardinality greater than one. As observed by Oikarinen and Janhunen (2008a), the heads of choice rules possessing multiple atoms can be freely split without affecting their semantics. When splitting such rules into $n$ different rules $\{a_i\} \leftarrow B, \sim C$ where $1 \leq i \leq n$, the only concern is the creation of $n$ copies of the rule body $B, \sim C$ which could reserve a quadratic space in the worst case. A new atom can be introduced to circumvent this. But the nature of proper disjunctive rules (1), the subject of study in this paper, is somewhat different. Unlike choice rules, disjunctive rules may interact through rule heads. In Example 2.3, the definition of $a$ depends on $b$ and vice versa. However, given a choice rule $\{a, b\} \leftarrow c$ for instance, the choices regarding $a$ and $b$ are independent of each other: if $c$ is true, both atoms can have any truth value. This is quite different from the interpretation of $a \vee b \leftarrow c$ which makes either $a$ or $b$ true given that $c$ is true. To grasp the interaction of $a$ and $b$ it is natural to have $b$ as an input to the definition of $a$ and, conversely, $a$ as input to that of $b$. As to be demonstrated in Section 7, shared rules can be rewritten so that input atoms are removed from the rule head but as a drawback of the rewriting technique, the compactness of the representation is partly lost. Therefore, we appreciate the extra flexibility provided by shared rules and interpret them to reflect the true nature of disjunctive rules.

In general, DLP-functions are composed according to the following principle:

---

6. Here and henceforth we make use of a tabular format to represent DLP-functions: the output signature is given on the top, the input signature at the bottom, and the rules are listed in between. Thus, the declaration of the hidden signature remains implicit.

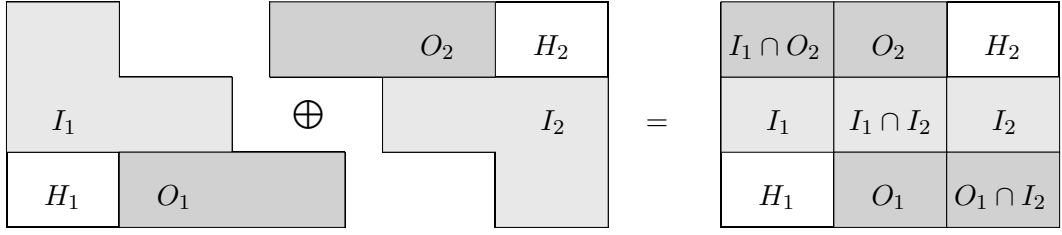

Figure 1: Treatment of signatures by the composition operator $\oplus$.

**Definition 2.4 (Composition)** *Let* $\Pi_1 = \langle R_1, I_1, O_1, H_1 \rangle$ *and* $\Pi_2 = \langle R_2, I_2, O_2, H_2 \rangle$ *be two DLP-functions that respect the input/output interfaces of each other. Then, the* composition *of* $\Pi_1$ *and* $\Pi_2$ *is defined and determined by*

$$\Pi_1 \oplus \Pi_2 = \langle R_1 \cup R_2, \ (I_1 \setminus O_2) \cup (I_2 \setminus O_1), \ O_1 \cup O_2, \ H_1 \cup H_2 \rangle. \tag{4}$$

The treatment of atom types under Definitions 2.2 and 2.4 is summarized in Figure 1. The two symmetric figures on the left-hand side illustrate the signatures of DLP-functions $\Pi_1 = \langle R_1, I_1, O_1, H_1 \rangle$ and $\Pi_2 = \langle R_2, I_2, O_2, H_2 \rangle$ subject to composition. Input signatures and output signatures are emphasized by light gray and dark gray shadings, respectively. The superposition of the two figures yields the diagram given on the right which represents the resulting nine categories of atoms. Only three of them may involve shared atoms that originate from both $\Pi_1$ and $\Pi_2$. The interface conditions introduced above should be intuitive to readers acquainted with the principles of object-oriented programming:

1. Although $\Pi_1$ and $\Pi_2$ must not share hidden atoms, they may share input atoms, i.e., $I_1 \cap I_2 \neq \emptyset$ is allowed. Output atoms are treated differently as $O_1 \cap O_2 = \emptyset$ is assumed.

2. An input atom of $\Pi_1$ becomes an output atom in $\Pi_1 \oplus \Pi_2$ if it appears as an output atom in $\Pi_2$, i.e., $\Pi_2$ provides the input for $\Pi_1$ in this setting. The input atoms of $\Pi_2$ are treated in a symmetric fashion.

3. The hidden atoms of $\Pi_1$ and $\Pi_2$ retain their status in $\Pi_1 \oplus \Pi_2$.

**Example 2.5** *Recall Example 2.3 in which we showed that DLP-functions* $\Pi_1$ *and* $\Pi_2$ *respect the input/output interfaces of each other. Thus, the composition of* $\Pi_1$ *and* $\Pi_2$ *is defined, and* $\Pi_1 \oplus \Pi_2$ *is* $\langle R_1 \cup R_2, I, O, H \rangle$ *where the set* $I$ *of input atoms is* $(\{a,c\} \setminus \{a\}) \cup (\{b,c\} \setminus \{b\}) = \{c\}$*, the set* $O$ *of output atoms is* $\{a\} \cup \{b\} = \{a,b\}$*, and the set* $H$ *of hidden atoms is* $\{d\} \cup \{e\} = \{d,e\}$*, i.e., using our tabular format to represent modules, we have*

| $\{b\}$ |
|---|
| $a \vee b \leftarrow c;$ $d \leftarrow a, \sim d$ |
| $\{a,c\}$ |

$\oplus$

| $\{a\}$ |
|---|
| $a \vee b \leftarrow c;$ $e \leftarrow a, \sim e$ |
| $\{b,c\}$ |

$=$

| $\{a,b\}$ |
|---|
| $a \vee b \leftarrow c;$ $d \leftarrow a, \sim d;$ $e \leftarrow a, \sim e$ |
| $\{c\}$ |

*The definitions of* $a$ *and* $b$ *in* $\Pi_1 \oplus \Pi_2$ *share the rule* $a \vee b \leftarrow c$. *Thanks to the flexibility of Definition 2.4, we are also able to split* $\Pi_1 \oplus \Pi_2$ *into its components whenever appropriate.*

Following previous approaches (Gelfond & Gabaldon, 1999; Oikarinen & Janhunen, 2008a), we define the signature $\mathrm{At}(\Pi)$ of a DLP-function $\Pi = \langle R, I, O, H \rangle$ as $I \cup O \cup H$.[7] For notational convenience, we distinguish the *visible* and *hidden parts* of $\mathrm{At}(\Pi)$ by setting $\mathrm{At_v}(\Pi) = I \cup O$ and $\mathrm{At_h}(\Pi) = H = \mathrm{At}(\Pi) \setminus \mathrm{At_v}(\Pi)$, respectively. Moreover, $\mathrm{At_i}(\Pi)$ and $\mathrm{At_o}(\Pi)$ are used to refer to the sets of input and output atoms of $\Pi$, respectively. These notations provide us a way to access the module interface when it is left implicit, e.g., to neglect the internal structure of modules. Lastly, for any set $S \subseteq \mathrm{At}(\Pi)$ of atoms, we denote the projections of $S$ on $\mathrm{At_i}(\Pi)$, $\mathrm{At_o}(\Pi)$, $\mathrm{At_v}(\Pi)$, and $\mathrm{At_h}(\Pi)$ by $S_\mathrm{i}$, $S_\mathrm{o}$, $S_\mathrm{v}$, and $S_\mathrm{h}$, respectively.

In formal terms, a DLP-function $\Pi = \langle R, I, O, H \rangle$ is designed to provide a mapping from subsets of $I$ to a set of subsets of $O \cup H$ in analogy to LP-functions formalized by Gelfond and Gabaldon (1999). However, the exact definition of this mapping is deferred until Section 3 where the semantics of DLP-functions will be anchored. In the sequel, the (syntactic) class of DLP-functions is denoted by $\mathcal{D}$. It is assumed, for the sake of simplicity, that $\mathcal{D}$ spans over a fixed (at most denumerable) signature $\mathrm{At}(\mathcal{D})$[8] so that $\mathrm{At}(\Pi) \subseteq \mathrm{At}(\mathcal{D})$ holds for each DLP-function $\Pi \in \mathcal{D}$. Given DLP-functions $\Pi_1$, $\Pi_2$, and $\Pi_3$ that pairwise respect the input/output interfaces of each other, it holds that

- $\Pi_1 \oplus \Pi_2 \in \mathcal{D}$ (closure),

- $\Pi_1 \oplus \varnothing = \varnothing \oplus \Pi_1 = \Pi_1$, for the empty DLP-function $\varnothing = \langle \emptyset, \emptyset, \emptyset, \emptyset \rangle$ (identity),

- $\Pi_1 \oplus \Pi_2 = \Pi_2 \oplus \Pi_1$ (commutativity), and

- $\Pi_1 \oplus (\Pi_2 \oplus \Pi_3) = (\Pi_1 \oplus \Pi_2) \oplus \Pi_3$ (associativity).

The theory of modules put forth by Oikarinen and Janhunen (2008a) is based on a more restrictive operator for program composition, viz. *the join* $\sqcup$. The idea behind this operator is to forbid positive dependencies between programs which is to be explicated next.

Technically speaking, we define the *positive dependency graph* $\mathrm{DG^+}(\Pi)$ of a DLP-function $\Pi = \langle R, I, O, H \rangle$ using only positive dependencies—following the definition by Ben-Eliyahu and Dechter (1994). However, we exclude input atoms from the graph as their definitions are external to $\Pi$ anyway. Thus, we let $\mathrm{DG^+}(\Pi) = \langle O \cup H, \leq_1 \rangle$ where $b \leq_1 a$ holds for a pair of atoms $a, b \in O \cup H$ if and only if there is a rule $A \leftarrow B, {\sim} C \in R$ such that $a \in A$ and $b \in B$. The reflexive and transitive closure of $\leq_1$ gives rise to the dependency relation $\leq$ over $\mathrm{At_o}(\Pi) \cup \mathrm{At_h}(\Pi)$. A *strongly connected component* (SCC) $S$ of the graph $\mathrm{DG^+}(\Pi)$ is a maximal set $S \subseteq \mathrm{At_o}(\Pi) \cup \mathrm{At_h}(\Pi)$ such that $b \leq a$ for every pair $a, b \in S$ of atoms. Given that $\Pi_1 \oplus \Pi_2$ is defined, we say that $\Pi_1$ and $\Pi_2$ are *mutually dependent* iff $\mathrm{DG^+}(\Pi_1 \oplus \Pi_2)$ has an SCC $S$ such that $S \cap \mathrm{At_o}(\Pi_1) \neq \emptyset$ and $S \cap \mathrm{At_o}(\Pi_2) \neq \emptyset$ (Oikarinen & Janhunen, 2008a), i.e., the component $S$ is shared by the DLP-functions $\Pi_1$ and $\Pi_2$ in this way. If $\Pi_1$ and $\Pi_2$ are not mutually dependent, we also call them *mutually independent*.

**Definition 2.6 (Joins)** *Given two DLP-functions $\Pi_1$ and $\Pi_2$, if the composition $\Pi_1 \oplus \Pi_2$ is defined and $\Pi_1$ and $\Pi_2$ are mutually independent, then the* join, $\Pi_1 \sqcup \Pi_2$, *of $\Pi_1$ and $\Pi_2$ is defined and it coincides with $\Pi_1 \oplus \Pi_2$.*

---

7. Consequently, the *length* of $\Pi$ in symbols, denoted by $\|\Pi\|$, gives an upper bound for $|\mathrm{At}(\Pi)|$ which is important when one considers the computational cost of translating programs (Janhunen, 2006).

8. In practice, this set could be the set of all identifiers (names for propositions or similar objects).

In case that $\Pi_1 \sqcup \Pi_2$ is defined, and thus $\Pi_1$ and $\Pi_2$ are mutually independent, exactly one of the following conditions holds for each SCC $S$ of $\mathrm{DG}^+(\Pi_1 \oplus \Pi_2)$:

$$S \subseteq \mathrm{At_o}(\Pi_1) \cup \mathrm{At_h}(\Pi_1); \text{ or} \tag{5}$$

$$S \subseteq \mathrm{At_o}(\Pi_2) \cup \mathrm{At_h}(\Pi_2). \tag{6}$$

**Example 2.7** *Recall the programs $\Pi_1$ and $\Pi_2$ from Example 2.5 for which we obtain the positive dependency graph $\mathrm{DG}^+(\Pi_1 \oplus \Pi_2) = \langle \{a,b,d,e\}, \{\langle a,d \rangle, \langle a,e \rangle\} \rangle$. Hence, the SCCs of the graph are simply singletons $\{a\}$, $\{b\}$, $\{d\}$, and $\{e\}$. Together with the observation that $\mathrm{At_o}(\Pi_1)$ and $\mathrm{At_o}(\Pi_2)$ are disjoint, we derive that $\Pi_1$ and $\Pi_2$ are not mutually dependent. Thus, the join $\Pi_1 \sqcup \Pi_2 = \Pi_1 \oplus \Pi_2$ is defined since the composition $\Pi_1 \oplus \Pi_2$ is defined on the basis of the analysis performed in Example 2.5.*

**Example 2.8** *As an example of two DLP-functions which have their composition defined yet which are ineligible for a join, consider the following situation:*

| $\{b\}$ |
|---|
| $a \vee b \leftarrow c;$ |
| $b \leftarrow a, \sim c$ |
| $\{a,c\}$ |

$\oplus$

| $\{a\}$ |
|---|
| $a \vee b \leftarrow c;$ |
| $a \leftarrow b, \sim c$ |
| $\{b,c\}$ |

$=$

| $\{a,b\}$ |
|---|
| $a \vee b \leftarrow c;$ |
| $a \leftarrow b, \sim c;$ |
| $b \leftarrow a, \sim c$ |
| $\{c\}$ |

*Here, the result of composition involves an SCC $S = \{a,b\}$ in the respective positive dependency graph, which has a non-empty intersection with the output signatures of the programs subject to composition. Hence, the respective join of the modules in question is not defined.*

## 3. Model Theory and Stable-Model Semantics

Having the syntax of DLP-functions defined, we now turn to their semantics. We proceed in three steps and introduce, correspondingly, three kinds of models, viz. classical models, minimal models, and, finally, stable models for each DLP-function. The last provide the intended semantics for a DLP-function whereas the first two serve as auxiliary concepts.

As usual, an *interpretation* for a DLP-function $\Pi$ is defined as an arbitrary subset of $\mathrm{At}(\Pi)$. Given a particular interpretation $M \subseteq \mathrm{At}(\Pi)$, an atom $a \in \mathrm{At}(\Pi)$ is *true* under $M$, denoted $M \models a$, iff $a \in M$, otherwise $a$ is *false* under $M$, denoted $M \not\models a$. For a negative literal $\sim a$, we define $M \models \sim a$ iff $M \not\models a$. A set $L$ of literals is *satisfied* by $M$, denoted by $M \models L$, iff $M \models l$, for every literal $l \in L$. We also define the disjunctive interpretation $\bigvee L$ of a set $L$ of literals: $M \models \bigvee L$ iff $M \models l$ for some literal $l \in L$.

To begin with, we cover DLP-functions with a pure classical semantics, which treats disjunctive rules as classical implications. It should be emphasized that classical models of a DLP-function $\Pi$ are specific interpretations as defined above and hence subsets of $\mathrm{At}(\Pi)$.

**Definition 3.1** *An interpretation $M \subseteq \mathrm{At}(\Pi)$ is a (classical) model of a DLP-function $\Pi = \langle R, I, O, H \rangle$, denoted $M \models \Pi$, iff $M \models R$, i.e., for every rule $A \leftarrow B, \sim C \in R$,*

$$M \models B \cup \sim C \text{ implies } M \models \bigvee A.$$

*The set of all classical models of $\Pi$ is denoted by* $\mathrm{CM}(\Pi)$.

Classical models provide an appropriate level of abstraction to address the role of input atoms in DLP-functions. Given a DLP-function $\Pi$ and an interpretation $M \subseteq \mathrm{At}(\Pi)$, the projection $M_\mathrm{i}$ can be viewed as the actual input for $\Pi$ which may (or may not) produce the respective output $M_\mathrm{o}$, depending on the semantics assigned to $\Pi$. The treatment of input atoms in the sequel will be based on *partial evaluation*: the idea is to pre-interpret input atoms appearing in $\Pi$ with respect to $M_\mathrm{i}$.

**Definition 3.2** *For a DLP-function* $\Pi = \langle R, I, O, H \rangle$ *and an actual input* $M_\mathrm{i} \subseteq I$ *for* $\Pi$, *the* instantiation *of* $\Pi$ *with respect to* $M_\mathrm{i}$, *denoted by* $\Pi/M_\mathrm{i}$, *is the quadruple* $\langle R', \emptyset, O, H \rangle$ *where* $R'$ *contains a reduced rule*

$$(A \setminus I) \leftarrow (B \setminus I), {\sim}(C \setminus I) \tag{7}$$

*for each rule* $A \leftarrow B, {\sim}C \in R$ *such that* $M_\mathrm{i} \models {\sim}A_\mathrm{i} \cup B_\mathrm{i} \cup {\sim}C_\mathrm{i}$.

**Example 3.3** *Consider the following DLP-function* $\Pi$:

| $\{a, b\}$ |
|---|
| $a \vee b \leftarrow {\sim}c;$ |
| $a \leftarrow c, {\sim}b;$ |
| $b \leftarrow c, {\sim}a$ |
| $\{c\}$ |

*For the actual input* $\{c\} \subseteq \mathrm{At}_\mathrm{i}(\Pi)$, *the reduct* $\Pi/\{c\}$ *is the DLP-function*

$$\langle \{a \leftarrow {\sim}b;\ b \leftarrow {\sim}a\}, \emptyset, \{a, b\}, \emptyset \rangle.$$

*On the other hand, with the actual input* $\emptyset \subseteq \mathrm{At}_\mathrm{i}(\Pi)$, *we obtain the reduct*

$$\Pi/\emptyset = \langle \{a \vee b\}, \emptyset, \{a, b\}, \emptyset \rangle.$$

The rules of form (7) are free of input atoms which indicates that the reduct $\Pi/M_\mathrm{i}$ is a DLP-function without input. Atoms in $\mathrm{At}_\mathrm{o}(\Pi) \cup \mathrm{At}_\mathrm{h}(\Pi)$ are not affected in $\Pi/M_\mathrm{i}$.

**Proposition 3.4** *Let* $\Pi$ *be a DLP-function and* $M \subseteq \mathrm{At}(\Pi)$ *an interpretation that defines an actual input* $M_\mathrm{i} \subseteq \mathrm{At}_\mathrm{i}(\Pi)$ *for* $\Pi$. *For all interpretations* $N \subseteq \mathrm{At}(\Pi)$ *such that* $N_\mathrm{i} = M_\mathrm{i}$,

$$N \models \Pi \iff N_\mathrm{o} \cup N_\mathrm{h} \models \Pi/M_\mathrm{i}.$$

*Proof.* Consider any $N \subseteq \mathrm{At}(\Pi)$ such that $N_\mathrm{i} = M_\mathrm{i}$.

($\Longrightarrow$) Suppose that $N \models \Pi$. Assume that $N_\mathrm{o} \cup N_\mathrm{h}$ does not satisfy (7) for some rule $A \leftarrow B, {\sim}C$ in $\Pi$. It follows that $M_\mathrm{i} \models {\sim}A_\mathrm{i} \cup B_\mathrm{i} \cup {\sim}C_\mathrm{i}$, and therefore $N_\mathrm{i} \models {\sim}A_\mathrm{i} \cup B_\mathrm{i} \cup {\sim}C_\mathrm{i}$. Thus, $N \not\models A \leftarrow B, {\sim}C$, a contradiction. It follows that $N_\mathrm{o} \cup N_\mathrm{h} \models \Pi/M_\mathrm{i}$.

($\Longleftarrow$) Let $N_\mathrm{o} \cup N_\mathrm{h} \models \Pi/M_\mathrm{i}$ hold. Assuming $N \not\models A \leftarrow B, {\sim}C$ for a rule of $\Pi$ implies $N \models B \cup {\sim}C$ and $N \not\models \bigvee A$. It follows that $N_\mathrm{i} \models {\sim}A_\mathrm{i} \cup B_\mathrm{i} \cup {\sim}C_\mathrm{i}$ and the corresponding rule (7) is included in $\Pi/M_\mathrm{i}$ as $N_\mathrm{i} = M_\mathrm{i}$. But this rule is not satisfied by $N_\mathrm{o} \cup N_\mathrm{h}$ since $N \not\models A \leftarrow B, \sim C$ implies $N_{\mathrm{o}} \cup N_{\mathrm{h}} \models (B \setminus I) \cup \sim(C \setminus I)$ and $N_{\mathrm{o}} \cup N_{\mathrm{h}} \not\models \bigvee(A \setminus I)$, a contradiction. Hence, we have that $N \models \Pi$. $\qquad\square$

Thus, the input reduction, as given in Definition 3.2, is fully compatible with classical semantics and we can characterize the semantic operator CM also in terms of the equation

$$\mathrm{CM}(\Pi) = \bigcup_{M_{\mathrm{i}} \subseteq \mathrm{At}_{\mathrm{i}}(\Pi)} \{M_{\mathrm{i}} \cup N \mid N \in \mathrm{CM}(\Pi/M_{\mathrm{i}})\}. \tag{8}$$

Recall that the models of any DLP-function $\Pi$ are subsets of $\mathrm{At}(\Pi)$. Hence, we have here that each $N \in \mathrm{CM}(\Pi/M_{\mathrm{i}})$ is a subset of $\mathrm{At}(\Pi/M_{\mathrm{i}})$ and thus $M_{\mathrm{i}} \cap N = \emptyset$ for each $M_{\mathrm{i}} \subseteq \mathrm{At}_{\mathrm{i}}(\Pi)$ since no atom from $\mathrm{At}_{\mathrm{i}}(\Pi)$ occurs in $\Pi/M_{\mathrm{i}}$ by definition.

Handling input atoms is slightly more complicated in the case of minimal models but the primitives of *parallel circumscription* (Lifschitz, 1985; McCarthy, 1986) provide us with a straightforward way to address them. The rough idea is to keep the interpretation of input atoms *fixed* while minimizing (i.e., falsifying) others as far as possible.

**Definition 3.5** *Let $\Pi = \langle R, I, O, H \rangle$ be a DLP-function. A model $M \subseteq \mathrm{At}(\Pi)$ of $\Pi$ is $I$-minimal iff there is no model $N$ of $\Pi$ such that $N_{\mathrm{i}} = M_{\mathrm{i}}$ and $N \subset M$.*

In the sequel, the set of $I$-minimal models of $\Pi = \langle R, I, O, H \rangle$ is denoted by $\mathrm{MM}(\Pi)$ and we treat input atoms by stipulating $I$-minimality of models. Using this idea, Proposition 3.4 lifts for minimal models given the fact that $\mathrm{At}_{\mathrm{i}}(\Pi/M_{\mathrm{i}}) = \emptyset$.

**Proposition 3.6** *Let $\Pi$ be a DLP-function and $M \subseteq \mathrm{At}(\Pi)$ an interpretation that defines an actual input $M_{\mathrm{i}} \subseteq \mathrm{At}_{\mathrm{i}}(\Pi)$ for $\Pi$. For all interpretations $N \subseteq \mathrm{At}(\Pi)$ such that $N_{\mathrm{i}} = M_{\mathrm{i}}$,*

$$N \in \mathrm{MM}(\Pi) \iff N_{\mathrm{o}} \cup N_{\mathrm{h}} \in \mathrm{MM}(\Pi/M_{\mathrm{i}}).$$

*Proof.* Consider any $N \subseteq \mathrm{At}(\Pi)$ such that $N_{\mathrm{i}} = M_{\mathrm{i}}$.

($\Longrightarrow$) Let $N \in \mathrm{MM}(\Pi)$. It follows by Proposition 3.4 that $N_{\mathrm{o}} \cup N_{\mathrm{h}} \models \Pi/M_{\mathrm{i}}$. Assume that $N_{\mathrm{o}} \cup N_{\mathrm{h}} \notin \mathrm{MM}(\Pi/M_{\mathrm{i}})$. Recall that $\mathrm{At}_{\mathrm{i}}(\Pi/M_{\mathrm{i}}) = \emptyset$. Thus, there is an interpretation $S \subset N_{\mathrm{o}} \cup N_{\mathrm{h}}$ such that $S \models \Pi/M_{\mathrm{i}}$. It follows by Proposition 3.4 that $N' \models \Pi$ for an interpretation $N' = M_{\mathrm{i}} \cup S$. But then $N'_{\mathrm{i}} = N_{\mathrm{i}}$ and $N' \subset N$ jointly contradict $N \in \mathrm{MM}(\Pi)$.

($\Longleftarrow$) Suppose that $N_{\mathrm{o}} \cup N_{\mathrm{h}} \in \mathrm{MM}(\Pi/M_{\mathrm{i}})$. So, $N_{\mathrm{o}} \cup N_{\mathrm{h}} \models \Pi/M_{\mathrm{i}}$, and $N \models \Pi$ follows by Proposition 3.4. Let us then assume that $N \notin \mathrm{MM}(\Pi)$, i.e., there is a model $N' \models \Pi$ with $N'_{\mathrm{i}} = N_{\mathrm{i}}$ and $N' \subset N$. Thus, we have $(N'_{\mathrm{o}} \cup N'_{\mathrm{h}}) \subset (N_{\mathrm{o}} \cup N_{\mathrm{h}})$, and since $N'_{\mathrm{i}} = N_{\mathrm{i}} = M_{\mathrm{i}}$ it follows that $N'_{\mathrm{o}} \cup N'_{\mathrm{h}} \models \Pi/M_{\mathrm{i}}$ by Proposition 3.4. Then, however, $N'_{\mathrm{o}} \cup N'_{\mathrm{h}} \models \Pi/M_{\mathrm{i}}$ is in contradiction with $N_{\mathrm{o}} \cup N_{\mathrm{h}} \in \mathrm{MM}(\Pi/M_{\mathrm{i}})$. $\qquad\square$

The set $\mathrm{MM}(\Pi)$ of $\mathrm{At}_{\mathrm{i}}(\Pi)$-minimal models is sufficient to determine the semantics of a *positive* DLP-function $\Pi$, i.e., whose rules are of the form $A \leftarrow B$. Recall that for such rules $A \setminus \mathrm{At}_{\mathrm{i}}(\Pi) \neq \emptyset$ holds whenever $A \neq \emptyset$. In order to cover arbitrary DLP-functions, we interpret negative body literals in the way proposed by Gelfond and Lifschitz (1991).

**Definition 3.7** *Given a DLP-function $\Pi = \langle R, I, O, H \rangle$ and an interpretation $M \subseteq \mathrm{At}(\Pi)$, the reduct of $\Pi$ with respect to $M$ is the positive DLP-function $\Pi^M = \langle R^M, I, O, H \rangle$ where*

$$R^M = \{A \leftarrow B \mid A \leftarrow B, \sim C \in R \text{ and } M \models \sim C\}. \tag{9}$$

**Definition 3.8** *An interpretation $M \subseteq \text{At}(\Pi)$ is a* stable model *of a DLP-function $\Pi$ with an input signature $\text{At}_i(\Pi)$ iff $M \in \text{MM}(\Pi^M)$, i.e., $M$ is an $\text{At}_i(\Pi)$-minimal model of $\Pi^M$.*

Hidden atoms play no special role in Definition 3.8. In contrast to this, they will affect possibilities for program decomposition, as to be presented in Section 6, and their status will be finally explicated when the notion of *modular equivalence* is introduced in Section 8. Definition 3.8 covers also the case of an *ordinary* disjunctive logic program, which is simply a DLP-function $\Pi = \langle R, \emptyset, O, \emptyset \rangle$: a model $M \subseteq \text{At}(\Pi) = O$ of $\Pi$ is stable iff $M$ is a minimal model of $R^M$. The definition of stable models gives rise to a semantic operator $\text{SM} : \mathcal{D} \to \mathbf{2}^{\mathbf{2}^{\text{At}(\mathcal{D})}}$ for DLP-functions:

$$\text{SM}(\Pi) = \{M \subseteq \text{At}(\Pi) \mid M \in \text{MM}(\Pi^M)\}. \tag{10}$$

Proposition 3.6 provides us a way to dismiss $\text{At}_i(\Pi)$-minimality in the definition of stable models if desirable. Given a stable model $M$ of $\Pi$, the projection $N = M_o \cup M_h$ is a minimal model of $(\Pi/M_i)^N$ and hence a stable model of $\Pi/M_i$. In other words, we have

$$(\Pi/M_i)^M = (\Pi/M_i)^{M_o \cup M_h} = \Pi^M/M_i.$$

Thus, we can derive the following result:

**Corollary 3.9** *For any DLP-function $\Pi$, we have*

$$\text{SM}(\Pi) = \{M \subseteq \text{At}(\Pi) \mid M_o \cup M_h \in \text{SM}(\Pi/M_i)\}.$$

**Example 3.10** *Recall the DLP-function $\Pi$ from Example 3.3, having no hidden atoms, given as follows:*

| $\{a, b\}$ |
|:---:|
| $a \vee b \leftarrow \sim c;$ |
| $a \leftarrow c, \sim b;$ |
| $b \leftarrow c, \sim a$ |
| $\{c\}$ |

*$\Pi$ has four stable models in total: $M_1 = \{a\}$, $M_2 = \{b\}$, $M_3 = \{a, c\}$, and $M_4 = \{b, c\}$, which are the $\{c\}$-minimal models of the respective reducts of $\Pi$:*

$$\begin{aligned}
\Pi^{M_1} &= \langle \{a \vee b \leftarrow; \ a \leftarrow c\}, \{c\}, \{a, b\}, \emptyset \rangle, \\
\Pi^{M_2} &= \langle \{a \vee b \leftarrow; \ b \leftarrow c\}, \{c\}, \{a, b\}, \emptyset \rangle, \\
\Pi^{M_3} &= \langle \{a \leftarrow c\}, \{c\}, \{a, b\}, \emptyset \rangle, \ and \\
\Pi^{M_4} &= \langle \{b \leftarrow c\}, \{c\}, \{a, b\}, \emptyset \rangle.
\end{aligned}$$

*Now, it is easy to verify that each $M_j$ is a $\{c\}$-minimal model of the reduct $\Pi^{M_j}$.*

*For illustrating Corollary 3.9, recall the reducts*

$$\begin{aligned}
\Pi/\{c\} &= \langle \{a \leftarrow \sim b; \ b \leftarrow \sim a\}, \emptyset, \{a, b\}, \emptyset \rangle \ and \\
\Pi/\emptyset &= \langle \{a \vee b\}, \emptyset, \{a, b\}, \emptyset \rangle.
\end{aligned}$$

*Then, we have that $\text{SM}(\Pi/\{c\}) = \{\{a\}, \{b\}\}$ and $\text{SM}(\Pi/\emptyset) = \{\{a\}, \{b\}\}$.*

An immediate observation is that we loose the general *antichain* property of stable models when input signatures are introduced. For instance, we have $M_1 \subset M_3$ and $M_2 \subset M_4$ in Example 3.10. However, since the interpretation of input atoms is fixed by the semantics, we perceive antichains *locally*, i.e., the set $\{N \in \mathrm{SM}(\Pi) \mid N_{\mathrm{i}} = M_{\mathrm{i}}\}$ of stable models forms an antichain, for each input $M_{\mathrm{i}} \subseteq \mathrm{At_i}(\Pi)$. In Example 3.10, the sets of stable models associated with actual inputs $\emptyset$ and $\{c\}$ are $\{M_1, M_2\}$ and $\{M_3, M_4\}$, respectively.

## 4. Characterizations using Classical Logic

It is well known how the set of stable models of an ordinary disjunctive logic program, i.e., a DLP-function $\Pi$ of the form $\langle R, \emptyset, O, \emptyset \rangle$, can be characterized via classical propositional logic, using the concepts of *completion* (Clark, 1978) and *loop formulas* (Lin & Zhao, 2004; Lee & Lifschitz, 2003). In this section, we generalize these concepts to arbitrary DLP-functions. To this end, the main concern is the role of input atoms and how to incorporate them into these concepts. Furthermore, we extend the *tightness* property of programs (Erdem & Lifschitz, 2003) to DLP-functions by introducing the notion of *I-tightness* in Section 4.2.

### 4.1 Program Completion and Loop Formulas

Given a DLP-function $\Pi$, a *loop* of $\Pi$ is any non-empty subset of a strongly connected component of the positive dependency graph $\mathrm{DG}^+(\Pi)$. Recall that $\mathrm{DG}^+(\Pi)$ has only the atoms of $\mathrm{At_o}(\Pi) \cup \mathrm{At_h}(\Pi)$ as its nodes. In particular, each singleton $\{a\}$ with $a \in \mathrm{At_o}(\Pi) \cup \mathrm{At_h}(\Pi)$ is thus a loop.

**Example 4.1** *Consider DLP-functions $\Pi_1$ and $\Pi_2$ defined as follows:*

$$
\Pi_1: \quad
\begin{array}{|c|}
\hline
\{b, c\} \\
\hline
a \vee c \leftarrow b; \\
b \leftarrow a \\
\hline
\{a\} \\
\hline
\end{array}
\qquad
\Pi_2: \quad
\begin{array}{|c|}
\hline
\{a, b\} \\
\hline
a \vee c \leftarrow b; \\
b \leftarrow a \\
\hline
\{c\} \\
\hline
\end{array}
$$

*Here, $\Pi_1$ has only singleton loops $\{b\}$ and $\{c\}$. In particular, $\{a, b\}$ is not a loop as it contains the input atom $a$. On the other hand, for $\Pi_2$ we have loops $\{a\}$, $\{b\}$, and $\{a, b\}$.*

In what follows, we use, for a set $S$ of propositional formulas (or atoms), $\neg S$ to denote a conjunction $\bigwedge_{s \in S} \neg s$ and $\bigvee S$ as a shorthand for $\bigvee_{s \in S} s$. Moreover, if appearing within a formula, a set $S$ is implicitly understood as a conjunction of its elements. For a DLP-function $\Pi$ and an atom $a \in \mathrm{At_o}(\Pi) \cup \mathrm{At_h}(\Pi)$, we define the set of *supporting formulas*

$$\mathrm{SuppF}(a, \Pi) = \{B \wedge \neg C \wedge \neg(A \setminus \{a\}) \mid A \leftarrow B, {\sim} C \in \Pi \text{ and } a \in A\}$$

and for a loop $L \subseteq \mathrm{At_o}(\Pi) \cup \mathrm{At_h}(\Pi)$ of $\Pi$, the set of *externally* supporting formulas

$$\mathrm{ESuppF}(L, \Pi) = \{B \wedge \neg C \wedge \neg(A \setminus L) \mid A \leftarrow B, {\sim} C \in \Pi, \ A \cap L \neq \emptyset, \text{ and } B \cap L = \emptyset\}.$$

Clark's completion procedure and (conjunctive) loop formulas can be generalized for DLP-functions in the following way:

**Definition 4.2** *For a DLP-function $\Pi$, the* completion *of $\Pi$ is the set of formulas*

$$\text{Comp}(\Pi) = \{B \wedge \neg C \to \bigvee A \mid A \leftarrow B, \sim C \in \Pi\} \cup$$
$$\{a \to \bigvee \text{SuppF}(a, \Pi) \mid a \in \text{At}_\text{o}(\Pi) \cup \text{At}_\text{h}(\Pi)\}$$

*and the set of* loop formulas *for* $\Pi$ *is*

$$\text{LF}(\Pi) = \{L \to \bigvee \text{ESuppF}(L, \Pi) \mid L \subseteq \text{At}_\text{o}(\Pi) \cup \text{At}_\text{h}(\Pi) \text{ is a loop of } \Pi\}.^9$$

Observe that in the case of $\text{At}_\text{i}(\Pi) = \emptyset$, i.e., $\text{At}_\text{o}(\Pi) \cup \text{At}_\text{h}(\Pi) = \text{At}(\Pi)$, the completion $\text{Comp}(\Pi)$ reduces to the definition provided by Lee and Lifschitz (2003) and the same holds for the set $\text{LF}(\Pi)$ of loop formulas. Generally speaking, the propositional theories $\text{Comp}(\Pi)$ and $\text{LF}(\Pi)$ characterize the set $\text{SM}(\Pi)$ of stable models in the following sense:

**Theorem 4.3** *For a DLP-function $\Pi$ and an interpretation $M \subseteq \text{At}(\Pi)$,*

$$M \in \text{SM}(\Pi) \text{ if and only if } M \models \text{Comp}(\Pi) \text{ and } M \models \text{LF}(\Pi).$$

*Proof.* We first relate the sets $\text{SuppF}(a, \Pi)$ and $\text{ESuppF}(L, \Pi)$, as introduced above for a DLP-function $\Pi$, with the respective sets of *complementary rules*

$$\text{SuppCR}(a, \Pi) = \{A \setminus \{a\} \leftarrow B, \sim C \mid A \leftarrow B, \sim C \in \Pi \text{ and } a \in A\} \text{ and}$$
$$\text{ESuppCR}(L, \Pi) = \{A \setminus L \leftarrow B, \sim C \mid A \leftarrow B, \sim C \in \Pi, \ A \cap L \neq \emptyset, \text{ and } B \cap L = \emptyset\}.$$

First, it is straightforward that, for each interpretation $M \subseteq \text{At}(\Pi)$, we have $M \models \text{Comp}(\Pi)$ iff jointly $M \models \Pi$ and for each $a \in M \cap (\text{At}_\text{o}(\Pi) \cup \text{At}_\text{h}(\Pi))$, $M \not\models \text{SuppCR}(a, \Pi)$. Quite similarly, it holds that $M \models \text{LF}(\Pi)$ iff, for each loop $L \subseteq M \cap (\text{At}_\text{o}(\Pi) \cup \text{At}_\text{h}(\Pi))$ of $\Pi$, $M \not\models \text{ESuppCR}(L, \Pi)$. On the other hand, by viewing $\text{SuppCR}(a, \Pi)$ and $\text{ESuppCR}(L, \Pi)$ as DLP-functions having the same signatures as $\Pi$, we can apply Proposition 3.4 in order to evaluate input atoms. Thus, we obtain the following relationships for each DLP-function $\Pi$, interpretation $M \subseteq \text{At}(\Pi)$, atom $a \in \text{At}_\text{o}(\Pi) \cup \text{At}_\text{h}(\Pi)$, and loop $L \subseteq \text{At}_\text{o}(\Pi) \cup \text{At}_\text{h}(\Pi)$ of $\Pi$:

1. $M \models \Pi$ iff $M_\text{o} \cup M_\text{h} \models \Pi / M_\text{i}$,

2. $M \models \text{SuppCR}(a, \Pi)$ iff $M_\text{o} \cup M_\text{h} \models \text{SuppCR}(a, \Pi / M_\text{i})$, and

3. $M \models \text{ESuppCR}(L, \Pi)$ iff $M_\text{o} \cup M_\text{h} \models \text{ESuppCR}(L, \Pi / M_\text{i})$.

Finally, recall that for each interpretation $M \subseteq \text{At}(\Pi)$, we have $\text{At}_\text{o}(\Pi) = \text{At}_\text{o}(\Pi / M_\text{i})$ and $\text{At}_\text{h}(\Pi) = \text{At}_\text{h}(\Pi / M_\text{i})$. Inspecting the definition of $\text{Comp}(\Pi)$ and $\text{LF}(\Pi)$ again, we can conclude for each interpretation $M \subseteq \text{At}(\Pi)$ that $M \models \text{Comp}(\Pi) \cup \text{LF}(\Pi)$ iff $M_\text{o} \cup M_\text{h} \models \text{Comp}(\Pi / M_\text{i}) \cup \text{LF}(\Pi / M_\text{i})$. In turn, we know that $M_\text{o} \cup M_\text{h} \models \text{Comp}(\Pi / M_\text{i}) \cup \text{LF}(\Pi / M_\text{i})$ iff $M_\text{o} \cup M_\text{h}$ is a stable model of the program $\Pi / M_\text{i}$ by the results of Lee and Lifschitz (2003); recall that $\Pi / M_\text{i}$ is an ordinary disjunctive program without any input atoms. Finally, we have $\text{SM}(\Pi) = \{M \subseteq \text{At}(\Pi) \mid M_\text{o} \cup M_\text{h} \in \text{SM}(\Pi / M_\text{i})\}$ by Corollary 3.9. This equality shows the claim. $\quad\square$

**Example 4.4** *Let us demonstrate the functioning of program completion and loop formulas on the DLP-functions from Example 4.1, i.e., on $\Pi_1 = \langle R, \{a\}, \{b, c\}, \emptyset \rangle$ and $\Pi_2 = \langle R, \{c\}, \{a, b\}, \emptyset \rangle$, where $R = \{a \vee c \leftarrow b; \ b \leftarrow a\}$. The completions are*

---

9. Although it may seem that the case of a singleton loop $L = \{a\}$ is somewhat redundant, this is not so, since some tautological rules such as $a \vee b \leftarrow a$ make a difference.

$$\mathrm{Comp}(\Pi_1) = \{b \to a \vee c, \ a \to b\} \cup \{b \to a, \ c \to b \wedge \neg a\} \ and$$
$$\mathrm{Comp}(\Pi_2) = \{b \to a \vee c, \ a \to b\} \cup \{b \to a, \ a \to b \wedge \neg c\}.$$

*Furthermore, the sets of loop formulas are*

$$\mathrm{LF}(\Pi_1) = \{b \to \bigvee \mathrm{ESuppF}(\{b\}, \Pi_1), \ c \to \bigvee \mathrm{ESuppF}(\{c\}, \Pi_1)\}$$
$$= \{b \to a, \ c \to b \wedge \neg a\} \ and$$
$$\mathrm{LF}(\Pi_2) = \{b \to \bigvee \mathrm{ESuppF}(\{b\}, \Pi_2), \ a \to \bigvee \mathrm{ESuppF}(\{a\}, \Pi_2),$$
$$a \wedge b \to \bigvee \mathrm{ESuppF}(\{a, b\}, \Pi_2)\}$$
$$= \{b \to a, \ a \to b \wedge \neg c, \ a \wedge b \to \bot\}.$$

*In the last formula, the occurrence of $\bot$ is in view of $\mathrm{ESuppF}(\{a, b\}, \Pi_2) = \emptyset$, which yields an empty disjunction $\bigvee \mathrm{ESuppF}(\{a, b\}, \Pi_2) = \bot$ as usual.*

*Computing the classical models of $\mathrm{Comp}(\Pi_1) \cup \mathrm{LF}(\Pi_1) = \mathrm{Comp}(\Pi_1)$ yields two such models, $M_1 = \{a, b\}$ and $M_2 = \emptyset$. One can check that these are indeed the stable models of $\Pi_1$ by recalling that $\mathrm{At_i}(\Pi_1) = \{a\}$. Thus, $M_1$ relates to an actual input $M_1 \cap \mathrm{At_i}(\Pi_1) = \{a\}$ whereas $M_2$ is based on $M_2 \cap \mathrm{At_i}(\Pi_1) = \emptyset$. On the other hand, the classical models of $\mathrm{Comp}(\Pi_2) \cup \mathrm{LF}(\Pi_2)$ are $M_1 = \{c\}$ and $M_2 = \emptyset$, which again relate to the two possible inputs over $\mathrm{At_i}(\Pi_2) = \{c\}$. Finally, we note that $\{a, b\}$ is also a model of $\mathrm{Comp}(\Pi_2)$ but ruled out by $\mathrm{LF}(\Pi_2)$.*

## 4.2 Tight DLP-functions

We now extend the well-known concept of *tightness* (Erdem & Lifschitz, 2003) to DLP-functions. This is of interest since we can exploit the fact that the positive dependency graph $\mathrm{DG^+}(\Pi)$ is reduced modulo input atoms. In other words, since the dependency graph $\mathrm{DG^+}(\Pi)$ has only the atoms of $\mathrm{At_o}(\Pi) \cup \mathrm{At_h}(\Pi)$ as its nodes, tightness for DLP-functions can be defined with respect to the input signature.

In the beginning of Section 4, loops were defined as arbitrary non-empty subsets of strongly connected components in $\mathrm{DG^+}(\Pi)$. Thus, if $\mathrm{DG^+}(\Pi)$ is acyclic then $\Pi$ has only singleton loops. However, the converse is not necessarily true, since, for a program $\Pi$ having only singleton loops, $\mathrm{DG^+}(\Pi)$ may have edges $\langle a, a \rangle$, i.e., cycles of length one.

**Definition 4.5** *A DLP-function $\Pi$ is $\mathrm{At_i}(\Pi)$-tight (or tight, for short), if the positive dependency graph $\mathrm{DG^+}(\Pi)$ is acyclic.*

**Example 4.6** *Recall DLP-functions $\Pi_1 = \langle R, \{a\}, \{b, c\}, \emptyset \rangle$ and $\Pi_2 = \langle R, \{c\}, \{a, b\}, \emptyset \rangle$ based on $R = \{a \vee c \leftarrow b; \ b \leftarrow a\}$ from Example 4.1. Here, $\Pi_1$ is $\{a\}$-tight since the potential non-singleton loop $\{a, b\}$ contains the input atom $a$. On the other hand, $\Pi_2$ is not $\{c\}$-tight. It is worth mentioning that the "ordinary" variant of $\Pi_1$, viz. DLP-function $\langle R, \emptyset, \{a, b, c\}, \emptyset \rangle$, is not $\emptyset$-tight—in particular, since $R$ is not tight in the usual sense.*

We note that the last observation, viz. that a DLP-function $\langle R, I, O, H \rangle$ may be $I$-tight although $R$ is not a tight program, relies on the use of disjunctions in the program. In fact, for DLP-functions $\langle R, I, O, H \rangle$, where $R$ is a set of normal rules of the form $a \leftarrow B, {\sim}C$, we have that a DLP-function $\Pi = \langle R, I, O, H \rangle$ is $I$-tight iff $R$ is tight. To verify this, note that the second item of Definition 2.1 implies that the head atom of a normal rule $a \leftarrow B, {\sim}C$ must not appear in $I$, and thus no loop of $\Pi$ may involve atoms from $I$.

We now show that the notion of tightness introduced in Definition 4.5 enables us to characterize the stable models of a DLP-function by the classical models of its completion. Since each "ordinary" program can be represented as a DLP-function, we thus properly generalize the well-known completion semantics (Clark, 1978). The following lemma is already sufficient for this result in view of Definition 4.2 and Theorem 4.3.

**Lemma 4.7** *For any tight DLP-function* $\Pi$, $\mathrm{LF}(\Pi) \subseteq \mathrm{Comp}(\Pi)$.

*Proof.* Recall that for each $a \in \mathrm{At_o}(\Pi) \cup \mathrm{At_h}(\Pi)$, $a \rightarrow \bigvee \mathrm{SuppF}(a, \Pi)$ is contained in $\mathrm{Comp}(\Pi)$. Moreover, since $\Pi$ is tight, $\Pi$ has only singleton loops, and thus $\mathrm{LF}(\Pi)$ contains only formulas $a \rightarrow \bigvee \mathrm{ESuppF}(\{a\}, \Pi)$, again for each $a \in \mathrm{At_o}(\Pi) \cup \mathrm{At_h}(\Pi)$. It remains to show that, for each atom $a$, $\mathrm{SuppF}(a, \Pi)$ is equivalent to $\mathrm{ESuppF}(\{a\}, \Pi)$ whenever the positive dependency graph is acyclic. We repeat the definition of $\mathrm{SuppF}(a, \Pi)$ and give the definition for $\mathrm{ESuppF}(L, \Pi)$, simplified for the case $L = \{a\}$:

$$\mathrm{SuppF}(a, \Pi) = \{B \wedge \neg C \wedge \neg(A \setminus \{a\}) \mid A \leftarrow B, \sim C \in \Pi \text{ and } a \in A\};$$
$$\mathrm{ESuppF}(\{a\}, \Pi) = \{B \wedge \neg C \wedge \neg(A \setminus \{a\}) \mid A \leftarrow B, \sim C \in \Pi, a \in A, \text{ and } B \cap \{a\} = \emptyset\}.$$

Now it is easy to see that for an acyclic dependency graph $\mathrm{DG}^+(\Pi)$, $a \in A$ implies $B \cap \{a\} = \emptyset$ for every rule $A \leftarrow B, \sim C \in \Pi$. Thus, we conclude that $\mathrm{SuppF}(a, \Pi) = \mathrm{ESuppF}(\{a\}, \Pi)$ holds for each $a \in \mathrm{At_o}(\Pi) \cup \mathrm{At_h}(\Pi)$. Hence, the claim follows. $\square$

**Example 4.8** *Recalling the DLP-function* $\Pi_1 = \langle R, \{a\}, \{b, c\}, \emptyset \rangle$ *from Example 4.4 with* $R = \{a \vee c \leftarrow b; \ b \leftarrow a\}$, *we obtain*

$$\mathrm{Comp}(\Pi_1) = \{b \rightarrow a \vee c, \ a \rightarrow b\} \cup \{b \rightarrow a, \ c \rightarrow b \wedge \neg a\} \ \text{and}$$
$$\mathrm{LF}(\Pi_1) = \{b \rightarrow \bigvee \mathrm{ESuppF}(\{b\}, \Pi_1), \ c \rightarrow \bigvee \mathrm{ESuppF}(\{c\}, \Pi_1)\}$$
$$= \{b \rightarrow a, \ c \rightarrow b \wedge \neg a\}.$$

*Now,* $\Pi_1$ *is tight and we observe that* $\mathrm{LF}(\Pi_1) \subseteq \mathrm{Comp}(\Pi_1)$ *as expected.*

The observations presented so far lead us to the following result:

**Theorem 4.9** *For a tight DLP-function* $\Pi$ *and an interpretation* $M \subseteq \mathrm{At}(\Pi)$,

$$M \in \mathrm{SM}(\Pi) \ \text{if and only if} \ M \models \mathrm{Comp}(\Pi).$$

In particular, this result is compatible with an existing characterization of stable models in the case of $\mathrm{At_i}(\Pi) = \emptyset$, i.e., if $\mathrm{At_o}(\Pi) \cup \mathrm{At_h}(\Pi) = \mathrm{At}(\Pi)$. Then, the notion of $\mathrm{At_i}(\Pi)$-tightness coincides with ordinary tightness, and the definition of the completion $\mathrm{Comp}(\Pi)$ reduces to the one provided by Lee and Lifschitz (2003).

## 5. Compositional Semantics

In what follows, our objective is to establish the main result of this paper, i.e., to show that stable-model semantics, as given by Definition 3.8, is fully compositional when larger DLP-functions $\Pi$ are formed as joins $\Pi_1 \sqcup \ldots \sqcup \Pi_n$ of DLP-functions. More precisely, the interconnection of $\mathrm{SM}(\Pi)$ and $\mathrm{SM}(\Pi_1), \ldots, \mathrm{SM}(\Pi_n)$ is explicated in Section 5.1. In analogy to Section 3, we follow a quite rigorous approach and consider such a relationship for classical models first, then for minimal models, and eventually cover the case of stable models which comprises our *module theorem*. Then, in Section 5.2, we use quantified Boolean formulas from the second level of polynomial hierarchy and their modular representation in terms of DLP-functions to illustrate the module theorem. Finally, we devote Section 5.3 to a comparison with the *splitting set theorem* proven by Lifschitz and Turner (1994).

## 5.1 Module Theorem

To begin with, we formalize the criteria for combining interpretations as well as models.

**Definition 5.1** *Given two DLP-functions $\Pi_1$ and $\Pi_2$, interpretations $M_1 \subseteq \mathrm{At}(\Pi_1)$ and $M_2 \subseteq \mathrm{At}(\Pi_2)$ are* mutually compatible *(with respect to $\Pi_1$ and $\Pi_2$), or just* compatible, *if*

$$M_1 \cap \mathrm{At_v}(\Pi_2) = M_2 \cap \mathrm{At_v}(\Pi_1). \tag{11}$$

According to (11), any two compatible interpretations $M_1$ and $M_2$ for $\Pi_1$ and $\Pi_2$, respectively, agree about the truth values of their *joint visible atoms* in $\mathrm{At_v}(\Pi_1) \cap \mathrm{At_v}(\Pi_2)$. A quick inspection of Figure 1 reveals the three cases that may arise when the join $\Pi = \Pi_1 \sqcup \Pi_2$ is defined and joint *output atoms* for $\Pi_1$ and $\Pi_2$ are thereafter disallowed: There may exist

1. joint input atoms in $\mathrm{At_i}(\Pi) = \mathrm{At_i}(\Pi_1) \cap \mathrm{At_i}(\Pi_2)$, or

2. atoms in $\mathrm{At_o}(\Pi_1) \cap \mathrm{At_i}(\Pi_2)$ that are output atoms in $\Pi_1$ and input atoms in $\Pi_2$, or

3. by symmetry, atoms in $\mathrm{At_i}(\Pi_1) \cap \mathrm{At_o}(\Pi_2)$.

Recall that according to Definition 2.6, atoms in the last two categories end up in $\mathrm{At_o}(\Pi)$ when $\Pi = \Pi_1 \sqcup \Pi_2$ is formed. Atoms in $\mathrm{At_v}(\Pi_1) \cap \mathrm{At_v}(\Pi_2)$ provide the basis to combine compatible interpretations for $\Pi_1$ and $\Pi_2$.

**Definition 5.2** *Let $\Pi_1$ and $\Pi_2$ be two DLP-functions such that $\Pi = \Pi_1 \sqcup \Pi_2$ is defined. Given any sets of interpretations $A_1 \subseteq 2^{\mathrm{At}(\Pi_1)}$ and $A_2 \subseteq 2^{\mathrm{At}(\Pi_2)}$, the* natural join *of $A_1$ and $A_2$ with respect to $\mathrm{At_v}(\Pi_1) \cap \mathrm{At_v}(\Pi_2)$, denoted by $A_1 \bowtie A_2$, is the set of interpretations*

$$A_1 \bowtie A_2 = \{M_1 \cup M_2 \mid M_1 \in A_1, \, M_2 \in A_2, \, \text{and } M_1 \text{ and } M_2 \text{ are compatible}\}. \tag{12}$$

Our first modularity result is formulated for DLP-functions under classical semantics as defined in Section 3. The combination of classical models is understood as in (12).

**Proposition 5.3** *For all* positive *DLP-functions $\Pi_1$ and $\Pi_2$ such that $\Pi_1 \oplus \Pi_2$ is defined,*

$$\mathrm{CM}(\Pi_1 \oplus \Pi_2) = \mathrm{CM}(\Pi_1) \bowtie \mathrm{CM}(\Pi_2). \tag{13}$$

*Proof.* Consider an interpretation $M \subseteq \mathrm{At}(\Pi_1 \oplus \Pi_2)$ and its projections $M_1 = M \cap \mathrm{At}(\Pi_1)$ and $M_2 = M \cap \mathrm{At}(\Pi_2)$ with respect to $\Pi_1 = \langle R_1, I_1, O_1, H_1 \rangle$ and $\Pi_2 = \langle R_2, I_2, O_2, H_2 \rangle$. It follows that $M_1$ and $M_2$ are compatible and $M = M_1 \cup M_2$ so that

$$
\begin{aligned}
M \in \mathrm{CM}(\Pi_1 \oplus \Pi_2) &\iff M \models R_1 \cup R_2 \\
&\iff M_1 \models R_1 \text{ and } M_2 \models R_2 \\
&\iff M_1 \in \mathrm{CM}(\Pi_1) \text{ and } M_2 \in \mathrm{CM}(\Pi_2) \\
&\iff M \in \mathrm{CM}(\Pi_1) \bowtie \mathrm{CM}(\Pi_2). \qquad \square
\end{aligned}
$$

Generalizing Proposition 5.3 for stable models of DLP-functions is much more elaborate. We will cover the case of positive DLP-functions under minimal models first. The proof of Theorem 5.5 exploits program completion, loop formulas, as well as the characterization of stable and minimal models from Section 4 as follows:

**Lemma 5.4** *For all DLP-functions $\Pi_1$ and $\Pi_2$ such that $\Pi_1 \sqcup \Pi_2$ is defined, the following conditions hold:*

$$
\begin{aligned}
\mathrm{Comp}(\Pi_1 \sqcup \Pi_2) &= \mathrm{Comp}(\Pi_1) \cup \mathrm{Comp}(\Pi_2); \quad (14) \\
\mathrm{LF}(\Pi_1 \sqcup \Pi_2) &= \mathrm{LF}(\Pi_1) \cup \mathrm{LF}(\Pi_2). \quad (15)
\end{aligned}
$$

*Proof.* We begin the proof by analyzing how formulas introduced by Clark's completion and loop formulas are related with joins of DLP-functions. To this end, we will now establish that the sets of formulas associated with $\Pi_1 \sqcup \Pi_2$ are directly obtained as unions of sets of formulas associated with $\Pi_1 = \langle R_1, I_1, O_1, H_1 \rangle$ and $\Pi_2 = \langle R_2, I_2, O_2, H_2 \rangle$: First, an implication $B \wedge \neg C \to \bigvee A$ belongs to $\mathrm{Comp}(\Pi_1 \sqcup \Pi_2)$ if and only if it belongs to $\mathrm{Comp}(\Pi_1)$, $\mathrm{Comp}(\Pi_2)$, or both in case of a shared rule. Second, let us consider any atom $a \in O \cup H$, where $O = O_1 \cup O_2$ and $H = H_1 \cup H_2$ are disjoint because $\Pi_1 \sqcup \Pi_2$ is defined. For the same reason, either $a \in O_1 \cup H_1$ or $a \in O_2 \cup H_2$, i.e., the atom $a$ is defined either by $\Pi_1$ or $\Pi_2$. Thus, we have either $\mathrm{Def}_{R_1}(a) = \mathrm{Def}_{R_1 \cup R_2}(a)$ or $\mathrm{Def}_{R_2}(a) = \mathrm{Def}_{R_1 \cup R_2}(a)$ by Definition 2.2, which implies that either $\mathrm{SuppF}(a, \Pi_1 \sqcup \Pi_2) = \mathrm{SuppF}(a, \Pi_1)$ or $\mathrm{SuppF}(a, \Pi_1 \sqcup \Pi_2) = \mathrm{SuppF}(a, \Pi_2)$. It follows that the implication $a \to \bigvee \mathrm{SuppF}(a, \Pi_1 \sqcup \Pi_2)$ is a member of $\mathrm{Comp}(\Pi_1 \sqcup \Pi_2)$ if and only if either (i) $a \to \bigvee \mathrm{SuppF}(a, \Pi_1)$ belongs to $\mathrm{Comp}(\Pi_1)$ or (ii) $a \to \bigvee \mathrm{SuppF}(a, \Pi_2)$ belongs to $\mathrm{Comp}(\Pi_2)$. Thus, we may conclude (14) for the completions involved.

Third, recall that each loop $L \subseteq \mathrm{At}(\Pi_1 \sqcup \Pi_2)$ of $\Pi_1 \sqcup \Pi_2$ is contained in some SCC $S$ of $\Pi_1 \sqcup \Pi_2$. It follows by (5), (6), and Definition 2.2 that either

1. $L \subseteq O_1 \cup H_1$ is a loop of $\Pi_1$ and $\mathrm{Def}_{R_1}(L) = \mathrm{Def}_{R_1 \cup R_2}(L)$, or

2. $L \subseteq O_2 \cup H_2$ is a loop of $\Pi_2$ and $\mathrm{Def}_{R_2}(L) = \mathrm{Def}_{R_1 \cup R_2}(L)$.

In the cases above, we have either $\mathrm{ESuppF}(L, \Pi_1 \sqcup \Pi_2) = \mathrm{ESuppF}(L, \Pi_1)$ or $\mathrm{ESuppF}(L, \Pi_1 \sqcup \Pi_2) = \mathrm{ESuppF}(L, \Pi_2)$. Thus, the respective loop formula $L \to \bigvee \mathrm{ESuppF}(L, \Pi_1 \sqcup \Pi_2)$ belongs to $\mathrm{LF}(\Pi_1 \sqcup \Pi_2)$ if and only if it is contained in $\mathrm{LF}(\Pi_1) \cup \mathrm{LF}(\Pi_2)$. $\square$

**Theorem 5.5** *For all positive DLP-functions $\Pi_1$ and $\Pi_2$ such that $\Pi_1 \sqcup \Pi_2$ is defined,*

$$
\mathrm{MM}(\Pi_1 \sqcup \Pi_2) = \mathrm{MM}(\Pi_1) \bowtie \mathrm{MM}(\Pi_2). \quad (16)
$$

*Proof.* Consider any $M \subseteq \mathrm{At}(\Pi_1 \sqcup \Pi_2)$ and the respective projections $M_1 = M \cap \mathrm{At}(\Pi_1)$ and $M_2 = M \cap \mathrm{At}(\Pi_2)$ which are compatible and, moreover, $M = M_1 \cup M_2$. We obtain the following chain of equivalences:

$$\begin{aligned}
M \in \mathrm{MM}(\Pi_1 \sqcup \Pi_2) &\iff M \models \mathrm{Comp}(\Pi_1 \sqcup \Pi_2) \text{ and } M \models \mathrm{LF}(\Pi_1 \sqcup \Pi_2) && [\text{Theorem 4.3}] \\
&\iff \begin{cases} M_1 \models \mathrm{Comp}(\Pi_1) \text{ and } M_1 \models \mathrm{LF}(\Pi_1) \\ M_2 \models \mathrm{Comp}(\Pi_2) \text{ and } M_2 \models \mathrm{LF}(\Pi_2) \end{cases} && [(14) \text{ and } (15)] \\
&\iff M_1 \in \mathrm{MM}(\Pi_1) \text{ and } M_2 \in \mathrm{MM}(\Pi_2) && [\text{Theorem 4.3}] \\
&\iff M \in \mathrm{MM}(\Pi_1) \bowtie \mathrm{MM}(\Pi_2). && [\text{Definition 5.2}]
\end{aligned}$$

$\square$

**Example 5.6** *Let us demonstrate the result of Theorem 5.5 in a practical setting using DLP-functions $\Pi_1$ and $\Pi_2$ as visualized below and their composition $\Pi = \langle R, \emptyset, \{a, b, c, d, e\}, \emptyset \rangle$.*

$$\Pi_1 : \begin{array}{|c|}
\hline
\{a,b,c\} \\
\hline
a \vee b \leftarrow; \\
a \leftarrow b; \\
b \leftarrow a; \\
a \leftarrow c; \\
c \vee d \vee e \leftarrow a,b \\
\hline
\{d,e\} \\
\hline
\end{array}
\qquad
\Pi_2 : \begin{array}{|c|}
\hline
\{d,e\} \\
\hline
d \leftarrow c; \\
e \leftarrow d; \\
d \leftarrow e; \\
c \vee d \vee e \leftarrow a,b \\
\hline
\{a,b,c\} \\
\hline
\end{array}$$

*The join $\Pi_1 \sqcup \Pi_2$ is defined because the SCCs of the composition $\Pi$ are $S_1 = \{a,b,c\}$ and $S_2 = \{d,e\}$. The $\mathrm{At_i}(\Pi_1)$-minimal models of $\Pi_1$ are $\{a,b,c\}$, $\{a,b,d\}$, $\{a,b,e\}$, and $\{a,b,d,e\}$. Likewise, calculating $\mathrm{MM}(\Pi_2)$, we get*

$$\mathrm{MM}(\Pi_2) = \{\emptyset, \{a\}, \{b\}, \{c,d,e\}, \{a,b,d,e\}, \{a,c,d,e\}, \{b,c,d,e\}, \{a,b,c,d,e\}\}.$$

*Hence, the only minimal model of $\Pi$ is $M = \{a,b,d,e\}$ and the compatibility condition underlying (16) correctly excludes $N = \{a,b,c,d,e\} \notin \mathrm{MM}(\Pi)$. Note that there is no support to $c$ being true in $\Pi_1$ when $d$ and $e$ are true. Accordingly, $c \vee d \vee e \leftarrow a, b$ is not active.*

We are now prepared to present our central result:

**Theorem 5.7 (Module Theorem)** *For all DLP-functions $\Pi_1$ and $\Pi_2$ such that $\Pi_1 \sqcup \Pi_2$ is defined,*

$$\mathrm{SM}(\Pi_1 \sqcup \Pi_2) = \mathrm{SM}(\Pi_1) \bowtie \mathrm{SM}(\Pi_2). \tag{17}$$

*Proof.* Again, we take an interpretation $M \subseteq \mathrm{At}(\Pi_1 \sqcup \Pi_2)$ and the respective compatible projections $M_1 = M \cap \mathrm{At}(\Pi_1)$ and $M_2 = M \cap \mathrm{At}(\Pi_2)$ into consideration. The proof of (17) can be based on (16) once a number of preliminary facts has been established:

1. The composition $\Pi_1^{M_1} \oplus \Pi_2^{M_2}$ is defined.

   Since $\Pi_1 \sqcup \Pi_2$ is defined, we know that $\Pi_1 \oplus \Pi_2$ is defined. This indicates that $\Pi_1$ and $\Pi_2$ respect the input/output interfaces of each other. The construction of $\Pi_1^{M_1}$ and $\Pi_2^{M_2}$ does not affect this property which implies that $\Pi_1^{M_1} \oplus \Pi_2^{M_2}$ is defined.

2. The join $\Pi_1^{M_1} \sqcup \Pi_2^{M_2}$ is defined.

   By the preceding item, the positive dependency graph $\mathrm{DG}^+(\Pi_1^{M_1} \oplus \Pi_2^{M_2})$ is defined. Let us assume that $\Pi_1^{M_1}$ and $\Pi_2^{M_2}$ are mutually dependent, i.e., there is an SCC $S$ of the graph above such that $S \cap \mathrm{At_o}(\Pi_1^{M_1}) \neq \emptyset$ and $S \cap \mathrm{At_o}(\Pi_2^{M_2}) \neq \emptyset$. Since the dependency graph has potentially fewer dependencies than the respective graph $\text{DG}^+(\Pi_1 \oplus \Pi_2)$ for $\Pi_1$ and $\Pi_2$, it follows that $S$ is contained in some SCC $S'$ of the latter. Since $\text{At}_\text{o}(\Pi_1^{M_1}) = \text{At}_\text{o}(\Pi_1)$ and $\text{At}_\text{o}(\Pi_2^{M_2}) = \text{At}_\text{o}(\Pi_2)$, we obtain $S' \cap \text{At}_\text{o}(\Pi_1) \neq \emptyset$ and $S' \cap \text{At}_\text{o}(\Pi_2) \neq \emptyset$. Thus, $\Pi_1$ and $\Pi_2$ are mutually dependent, a contradiction.

3. The reduct $(\Pi_1 \sqcup \Pi_2)^M$ coincides with $\Pi_1^{M_1} \sqcup \Pi_2^{M_2}$.

   A rule $A \leftarrow B$ belongs to $(\Pi_1 \sqcup \Pi_2)^M$ if and only if there is a rule $A \leftarrow B, \sim C$ in $\Pi_1$, $\Pi_2$, or both such that $C \cap M = \emptyset$. Equivalently, there is a rule $A \leftarrow B, \sim C$ in $\Pi_1$ such that $C \cap M_1 = \emptyset$, or there is a rule $A \leftarrow B, \sim C$ in $\Pi_2$ such that $C \cap M_2 = \emptyset$, i.e., $A \leftarrow B \in \Pi_1^{M_1}$ or $A \leftarrow B \in \Pi_2^{M_2}$.

We therefore get the following chain of equivalences:

$$
\begin{aligned}
M \in \text{SM}(\Pi_1 \sqcup \Pi_2) &\iff M \in \text{MM}((\Pi_1 \sqcup \Pi_2)^M) && \text{[Definition 3.8]} \\
&\iff M \in \text{MM}(\Pi_1^{M_1} \sqcup \Pi_2^{M_2}) && \text{[Item 3 above]} \\
&\iff M \in \text{MM}(\Pi_1^{M_1}) \bowtie \text{MM}(\Pi_2^{M_2}) && \text{[Theorem 5.5]} \\
&\iff M_1 \in \text{MM}(\Pi_1^{M_1}) \text{ and } M_2 \in \text{MM}(\Pi_2^{M_2}) && \text{[Definition 5.2]} \\
&\iff M_1 \in \text{SM}(\Pi_1) \text{ and } M_2 \in \text{SM}(\Pi_2) && \text{[Definition 3.8]} \\
&\iff M \in \text{SM}(\Pi_1) \bowtie \text{SM}(\Pi_2). && \text{[Definition 5.2]} \qquad \square
\end{aligned}
$$

The moral of Theorem 5.7 and Definition 2.6 is that stable semantics supports modularization as long as positively interdependent atoms are enforced in the same module.

**Example 5.8** *Let $\Pi_1$ and $\Pi_2$ be DLP-functions as defined below and $\Pi = \Pi_1 \sqcup \Pi_2$ their join (which is clearly defined):*

$$
\begin{array}{|c|}
\hline
\{b\} \\
\hline
a \vee b \leftarrow; \\
b \vee c \leftarrow \\
\hline
\{a,c\} \\
\hline
\end{array}
\quad \sqcup \quad
\begin{array}{|c|}
\hline
\{c\} \\
\hline
a \vee c \leftarrow; \\
b \vee c \leftarrow \\
\hline
\{a,b\} \\
\hline
\end{array}
\quad = \quad
\begin{array}{|c|}
\hline
\{b,c\} \\
\hline
a \vee b \leftarrow; \\
a \vee c \leftarrow; \\
b \vee c \leftarrow \\
\hline
\{a\} \\
\hline
\end{array}
$$

*It is straightforward to verify that $\text{SM}(\Pi_1) = \{\{b\}, \{a,b\}, \{a,c\}, \{b,c\}\}$ and $\text{SM}(\Pi_2) = \{\{c\}, \{a,b\}, \{a,c\}, \{b,c\}\}$. Since $\text{At}_\text{v}(\Pi_1) \cap \text{At}_\text{v}(\Pi_2) = \{a,b,c\}$, we obtain*

$$\text{SM}(\Pi_1) \bowtie \text{SM}(\Pi_2) = \text{SM}(\Pi_1) \cap \text{SM}(\Pi_2) = \{\{a,b\}, \{a,c\}, \{b,c\}\}.$$

*A simple cross-check confirms that $\text{SM}(\Pi)$ is indeed given by this set.*

**Example 5.9** *Consider the DLP-functions $\Pi_1$ and $\Pi_2$ from Example 2.8. Then, $\text{SM}(\Pi_1) = \{\emptyset, \{a,b\}, \{b,c\}\}$ and $\text{SM}(\Pi_2) = \{\emptyset, \{a,b\}, \{a,c\}\}$. As shown in Example 2.8, the join of $\Pi_1$ and $\Pi_2$ is undefined. Thus, Theorem 5.7 is not applicable. Concerning the composition $\Pi_1 \oplus \Pi_2$, we note that $\text{SM}(\Pi_1 \oplus \Pi_2) = \{\emptyset, \{a,c\}, \{b,c\}\} \neq \{\emptyset, \{a,b\}\} = \text{SM}(\Pi_1) \bowtie \text{SM}(\Pi_2)$.*

Theorem 5.7 can be easily extended for DLP-functions consisting of more than two modules. In view of this, we say that a finite sequence $M_1, \ldots, M_n$ of stable models for modules $\Pi_1, \ldots, \Pi_n$, respectively, is *compatible*, iff $M_i$ and $M_j$ are pairwise compatible, for all $1 \leq i, j \leq n$. This property guarantees that each $M_i$ can be recovered from the union $M = \bigcup_{i=1}^{n} M_i$ by taking the respective projection $M \cap \text{At}(\Pi_i) = M_i$.

**Corollary 5.10** *Let $\Pi_1, \ldots, \Pi_n$ be a sequence of DLP-functions such that the join $\Pi_1 \sqcup \cdots \sqcup \Pi_n$ is defined. Then,*

$$\mathrm{SM}(\Pi_1 \sqcup \cdots \sqcup \Pi_n) = \mathrm{SM}(\Pi_1) \bowtie \cdots \bowtie \mathrm{SM}(\Pi_n). \tag{18}$$

**Example 5.11** *The following example simply extends Example 5.8:*

$$
\begin{array}{|c|}
\hline
\{b\} \\
\hline
a \vee b \leftarrow; \\
b \vee c \leftarrow \\
\hline
\{a, c\} \\
\hline
\end{array}
\;\sqcup\;
\begin{array}{|c|}
\hline
\{c\} \\
\hline
a \vee c \leftarrow; \\
b \vee c \leftarrow \\
\hline
\{a, b\} \\
\hline
\end{array}
\;\sqcup\;
\begin{array}{|c|}
\hline
\{a\} \\
\hline
a \vee b \leftarrow; \\
a \vee c \leftarrow \\
\hline
\{b, c\} \\
\hline
\end{array}
\;=\;
\begin{array}{|c|}
\hline
\{a, b, c\} \\
\hline
a \vee b \leftarrow; \\
a \vee c \leftarrow; \\
b \vee c \leftarrow \\
\hline
\emptyset \\
\hline
\end{array}
$$

*Now we have* $\mathrm{SM}(\Pi_1) = \{\{b\}, \{a, b\}, \{a, c\}, \{b, c\}\}$, $\mathrm{SM}(\Pi_2) = \{\{c\}, \{a, b\}, \{a, c\}, \{b, c\}\}$, *and* $\mathrm{SM}(\Pi_3) = \{\{a\}, \{a, b\}, \{a, c\}, \{b, c\}\}$. *Thus, we learn from Corollary 5.10 that*

$$\mathrm{SM}(\Pi_1 \sqcup \Pi_2 \sqcup \Pi_3) = \mathrm{SM}(\Pi_1) \bowtie \mathrm{SM}(\Pi_2) \bowtie \mathrm{SM}(\Pi_3) = \{\{a, b\}, \{a, c\}, \{b, c\}\}.$$

### 5.2 Modular Representation of Quantified Boolean Formulas

Our next objective is to illustrate the theory developed so far in terms of a more extensive example. To this end, we consider the pair of DLP-functions $\Pi_n^{\mathrm{sat}}$ and $\Pi_n^{\mathrm{unsat}}$ as depicted in Figure 2. Their purpose is the evaluation of *quantified Boolean formulas* (QBFs) of the form

$$\exists X \forall Y \bigvee_{i=1}^n (\neg A_i \wedge B_i \wedge \neg C_i \wedge D_i), \tag{19}$$

where each $A_j$, $B_j$, $C_j$, and $D_j$ is a set of Boolean variables, and the parameter $n$ gives the number of disjuncts in the *matrix* which is a Boolean formula $\phi$ in disjunctive normal form (DNF).[10] Without loss of generality, we may assume that $X = \bigcup_{i=1}^n (A_i \cup B_i)$, $Y = \bigcup_{i=1}^n (C_i \cup D_i)$, and $X \cap Y = \emptyset$ hold for the sets $X$ and $Y$ of Boolean variables in (19).

It is important to point out that in general the evaluation of QBFs of the form (19) constitutes a $\Sigma_2^p$-complete decision problem which perfectly matches the complexity of checking the existence of stable models for a disjunctive program. Given this completeness property, it follows that in principle any decision problem in $\Sigma_2^p$ can be turned into a QBF of the form (19), albeit more direct representations can be obtained for particular problem domains. In this respect, let us address three specific domains prior to detailing the generic approach.

1. The *strategic companies* domain is identified by Leone et al. (2006) as one of the first practical domains involving decision problems on the second level of the polynomial-time hierarchy and solved using ASP techniques. The simplified encoding provided by Koch, Leone, and Pfeifer (2003) is based on two kinds of disjunctive rules:

$$\mathsf{strat}(x_1) \vee \mathsf{strat}(x_2) \vee \mathsf{strat}(x_3) \vee \mathsf{strat}(x_4) \leftarrow \mathsf{prod}(y, x_1, x_2, x_3, x_4), \tag{20}$$

$$\mathsf{strat}(x) \leftarrow \mathsf{ctrl}(x, x_1, x_2, x_3, x_4), \mathsf{strat}(x_1), \mathsf{strat}(x_2), \mathsf{strat}(x_3), \mathsf{strat}(x_4), \tag{21}$$

---

10. Also, recall the shorthands $\neg S = \bigwedge_{s \in S} \neg s$ and $S = \bigwedge_{s \in S} s$ introduced right after Example 4.1.

Function $\Pi_n^{\text{sat}}$:

| $X$ | |
|---|---|
| For $1 \leq i \leq n$ and $x \in A_i$: | $\leftarrow x, \text{act}(i);$ |
| For $1 \leq i \leq n$ and $x \in B_i$: | $x \leftarrow \text{act}(i);$ |
| For $1 \leq i \leq n$: | $A_i \leftarrow B_i, \sim\!\text{act}(i)$ |
| $\{\text{act}(1), \ldots, \text{act}(n)\}$ | |

Function $\Pi_n^{\text{unsat}}$:

| $\emptyset$ | |
|---|---|
| For $1 \leq i \leq n$: | $C_i \cup \{u\} \leftarrow D_i, \text{act}(i);$ |
| For $y \in Y$: | $y \leftarrow u;$ |
| | $u \leftarrow \sim\!u$ |
| $\{\text{act}(1), \ldots, \text{act}(n)\}$ | |

Figure 2: DLP-functions $\Pi_n^{\text{sat}}$ and $\Pi_n^{\text{unsat}}$ for the evaluation of a quantified Boolean formula $\forall X \exists Y \phi$ having a matrix $\phi = \bigvee_{i=1}^n (\neg A_i \wedge B_i \wedge \neg C_i \wedge D_i)$.

where predicates $\text{strat}(x)$, $\text{prod}(y, x_1, x_2, x_3, x_4)$, and $\text{ctrl}(x, x_1, x_2, x_3, x_4)$, respectively, denote that a company $x$ is strategic, a product $y$ is produced by companies $x_1, \ldots, x_4$, and a company $x$ is controlled by companies $x_1, \ldots, x_4$. Obviously, instances of the predicate $\text{strat}$ arising from the rules of the forms (20) and (21) create positive dependencies in such a program $\Pi$. The resulting SCCs can be used to split the program into modules $\Pi_1, \ldots, \Pi_n$ so that $\Pi = \Pi_1 \sqcup \ldots \sqcup \Pi_n$ is defined. By Theorem 5.7, the status of a specific company $x$ can be decided using the module $\Pi_i$ which defines $\text{strat}(x)$ rather than the entire encoding $\Pi$.

2. The *model-based diagnosis* of digital circuitry provides another interesting application area. Quite recently, Oikarinen and Janhunen (2008b) presented an efficient encoding of *prioritized circumscription* as a disjunctive program (and thus, as a special case, of *parallel circumscription* as well)—enabling a concise representation of minimal diagnoses in the sense of Reiter (1987). The resulting disjunctive rules involve *head-cycles* (see Section 7 for details) which typically pre-empt a polynomial-time translation into a computationally easier *normal* logic program. This observation suggests completeness on the second level of the polynomial-time hierarchy although we are not aware of an exact hardness result. The correctness proof of the encoding exploits two modules and the module theorem.

3. Finally, let us mention that Gebser, Schaub, Thiele, Usadel, and Veber (2008b) identify *minimal inconsistent cores* in large biological networks with disjunctive programs. The decision problem in question is $D^{\text{p}}$-complete which also indicates the appropriateness of disjunctive logic programs for the representation of this domain. Since any $D^{\text{p}}$-complete decision problem can be described as an independent combination of an NP-complete decision problem $P_1$ and a coNP-complete decision problem $P_2$, we foresee a representation in the form of a join $\Pi^{\text{sat}} \sqcup \Pi^{\text{unsat}}$, where $\Pi^{\text{sat}}$ has a stable model iff $P_1$ has a succinct certificate, and $\Pi^{\text{unsat}}$ has a unique stable model iff $P_2$ has no succinct certificates. The required DLPs can be worked out via reductions into propositional (un)satisfiability. In particular, the test for unsatisfiability can be realized in analogy to $\Pi_n^{\mathrm{unsat}}$ analyzed below.

| $\{x_1, x_2\}$ | |
|---|---|
| $x_1 \leftarrow \mathsf{act}(1);$ | $\leftarrow x_1, \sim\mathsf{act}(1);$ |
| $\leftarrow x_2, \mathsf{act}(2);$ | $x_2 \leftarrow \sim\mathsf{act}(2);$ |
| $x_1 \leftarrow \mathsf{act}(3);$ | $\leftarrow x_1, \sim\mathsf{act}(3);$ |
| $x_1 \leftarrow \mathsf{act}(4);$ | $\leftarrow x_2, \mathsf{act}(4);$ |
| $x_2 \leftarrow x_1, \sim\mathsf{act}(4)$ | |
| $\{\mathsf{act}(1), \mathsf{act}(2), \mathsf{act}(3), \mathsf{act}(4)\}$ | |

| $\emptyset$ |
|---|
| $u \leftarrow y_1, y_2, \mathsf{act}(1);$ |
| $u \vee y_2 \leftarrow y_1, \mathsf{act}(2);$ |
| $u \vee y_1 \leftarrow y_2, \mathsf{act}(3);$ |
| $u \vee y_1 \vee y_2 \leftarrow \mathsf{act}(4);$ |
| $y_1 \leftarrow u; \ y_2 \leftarrow u; \ u \leftarrow \sim u$ |
| $\{\mathsf{act}(1), \mathsf{act}(2), \mathsf{act}(3), \mathsf{act}(4)\}$ |

Figure 3: Particular instances of $\Pi_4^{\mathrm{sat}}$ and $\Pi_4^{\mathrm{unsat}}$.

In the general case, we use Boolean variables and propositional atoms interchangeably in order to describe how the validity problem of (19) is captured by DLP-functions from Figure 2. The design of $\Pi_n^{\mathrm{sat}}$ and $\Pi_n^{\mathrm{unsat}}$ is based on the *explanatory* approach from Janhunen et al. (2006), where (19) is equivalently viewed as a formula $\exists X \neg \exists Y \neg \phi$ having the matrix $\neg \phi$ in *conjunctive normal form* (CNF). A *clause*[11] $A_i \vee \neg B_i \vee C_i \vee \neg D_i$ in $\neg \phi$ is *active* whenever $A_i \vee \neg B_i$ is false and the truth of the clause becomes dependent on $C_i \vee \neg D_i$; or to put it dually, $\neg A_i \wedge B_i$ is true and the truth of $\neg A_i \wedge B_i \wedge \neg C_i \wedge D_i$ depends on $\neg C_i \wedge D_i$. The validity of the formula $\exists X \neg \exists Y \neg \phi$ is captured as follows: Given an input interpretation $M_{\mathrm{i}} \subseteq \{\mathsf{act}(1), \ldots, \mathsf{act}(n)\}$, the upper DLP-function $\Pi_n^{\mathrm{sat}}$ from Figure 2 tries to *explain* the activation statuses of the clauses in $\neg \phi$ by checking that the respective theory $\{\neg A_i \wedge B_i \mid \mathsf{act}(i) \in M_{\mathrm{i}}\} \cup \{A_i \vee \neg B_i \mid \mathsf{act}(i) \notin M_{\mathrm{i}}\}$ is satisfiable. The lower DLP-function, $\Pi_n^{\mathrm{unsat}}$, plays the role of a coNP-*oracle*: it captures a test for the theory $\{C_i \vee \neg D_i \mid \mathsf{act}(i) \in M_{\mathrm{i}}\}$ being unsatisfiable. The correctness of the representation provided by these DLP-functions will be addressed soon, but it is enough to understand their syntax and intuitive meaning for the moment. A concrete QBF instance is evaluated as follows.

**Example 5.12** *Consider DLP-functions $\Pi_n^{\mathrm{sat}}$ and $\Pi_n^{\mathrm{unsat}}$ from Figure 2 in the case of QBF*

$$\exists x_1 \exists x_2 \forall y_1 \forall y_2 [(x_1 \wedge y_1 \wedge y_2) \vee (\neg x_2 \wedge y_1 \wedge \neg y_2) \vee (x_1 \wedge \neg y_1 \wedge y_2) \vee (x_1 \wedge \neg x_2 \wedge \neg y_1 \wedge \neg y_2)]. \quad (22)$$

*Thus, the parameter for this instance is $n = 4$, and the input signature is $\{\mathsf{act}(1), \ldots, \mathsf{act}(4)\}$ for both $\Pi_4^{\mathrm{sat}}$ and $\Pi_4^{\mathrm{unsat}}$, as illustrated in Figure 3. The output signature of the former DLP-function is $\{x_1, x_2\}$ and all other atoms, i.e., $y_1$, $y_2$, and $u$, remain hidden in the latter. The joint input signature is used to specify the active part of the matrix in (22). The DLP-function $\Pi_4^{\mathrm{sat}}$ provides an explanation, i.e., an assignment to variables $x_1$ and $x_2$ as its output, whereas $\Pi_4^{\mathrm{unsat}}$ is only responsible for the respective unsatisfiability check. As regards the validity of the QBF given in (22), the input interpretation $\{\mathsf{act}(1), \mathsf{act}(2), \mathsf{act}(3), \mathsf{act}(4)\}$ yields a positive answer. The respective explanation, i.e., the output interpretation found by $\Pi_4^{\mathrm{sat}}$, is $\{x_1\}$. It is easy to check that when $x_1$ is true and $x_2$ is false then the remainder of the matrix is true whatever values are assigned to $y_1$ and $y_2$. Hence, the QBF (22) is valid.*

---

11. For the purposes of this section, we interpret disjunctions $A \vee \neg B$ of sets $A$ and $\neg B = \{\neg b \mid b \in B\}$ of positive and negative literals, respectively, as disjunctions of their elements.

As regards the general DLP-functions $\Pi_n^{\text{sat}}$ and $\Pi_n^{\text{unsat}}$ in Figure 2, they have identical input signatures, only $\Pi_n^{\text{sat}}$ has output atoms, and the hidden atoms of $\Pi_n^{\text{unsat}}$ are fully respected. Hence, the composition $\Pi_n^{\text{sat}} \oplus \Pi_n^{\text{unsat}}$ is defined. Moreover, the atoms appearing in rules that involve positive dependencies belong to disjoint sets $X$ and $Y \cup \{u\}$. It is therefore clear that $\text{DG}^+(\Pi_n^{\text{sat}} \oplus \Pi_n^{\text{unsat}})$ cannot have an SCC $S$ such that $S \cap X \neq \emptyset$ and $S \cap (Y \cup \{u\}) \neq \emptyset$. This implies that $\Pi_n^{\text{sat}} \sqcup \Pi_n^{\text{unsat}}$ is defined regardless of the QBF (19) in question. Let us exploit this fact in the context of specific DLP-functions of Example 5.12.

**Example 5.13**  *There are four stable models for the DLP-function $\Pi_4^{\text{sat}}$:*

$$\{\text{act}(1), \text{act}(2), \text{act}(3), \text{act}(4), x_1\}, \ \{\text{act}(1), \text{act}(3), x_1, x_2\}, \ \{\text{act}(2)\}, \ \text{and } \{x_2\},$$

*listed in decreasing level of activation. On the other hand, the DLP-function $\Pi_4^{\text{unsat}}$ has a unique stable model $\{\text{act}(1), \text{act}(2), \text{act}(3), \text{act}(4), y_1, y_2, u\}$, i.e., the interpretation $\{y_1, y_2, u\}$ is the unique stable model of $\Pi_4^{\text{unsat}}/\{\text{act}(1), \text{act}(2), \text{act}(3), \text{act}(4)\}$ where the set of rules is given by*

$$\{ \ u \leftarrow y_1, y_2; \ \ u \vee y_2 \leftarrow y_1; \ \ u \vee y_1 \leftarrow y_2; \ \ u \vee y_1 \vee y_2; \ \ y_1 \leftarrow u; \ \ y_2 \leftarrow u; \ \ u \leftarrow \sim u \ \},$$

*and $\Pi_4^{\text{unsat}}/M_i$ has no stable models for any other input interpretation $M_i$. Moreover, we may apply the module theorem to calculate $\text{SM}(\Pi_4^{\text{sat}} \sqcup \Pi_4^{\text{unsat}})$ by combining compatible pairs of models. There is only one such pair:*

$$\{\text{act}(1), \text{act}(2), \text{act}(3), \text{act}(4), x_1\} \in \text{SM}(\Pi_4^{\text{sat}}) \ and$$
$$\{\text{act}(1), \text{act}(2), \text{act}(3), \text{act}(4), y_1, y_2, u\} \in \text{SM}(\Pi_4^{\text{unsat}}).$$

*Thus, $\{\text{act}(1), \text{act}(2), \text{act}(3), \text{act}(4), x_1, y_1, y_2, u\}$ is the unique stable model of the join $\Pi_4^{\text{sat}} \sqcup \Pi_4^{\text{unsat}}$. Since $\text{SM}(\Pi_4^{\text{sat}} \sqcup \Pi_4^{\text{unsat}})$ is non-empty, we conclude that (22) is indeed valid.*

It is natural to ask what can be stated about the stable models of the general DLP-functions $\Pi_n^{\text{unsat}}$ and $\Pi_n^{\text{sat}}$ associated with the QBF $\exists X \forall Y \phi$ given in (19). If $M$ is a stable model of $\Pi_n^{\text{sat}}$, then the respective projection $M_X = X \cap M$ *determines* $M$, i.e., it holds for all $1 \leq i \leq n$ in the matrix $\phi$ that $\text{act}(i) \in M$ if and only if $M_X \models \neg A_i \wedge B_i$. Moreover, the model $M_X$ is minimal in the sense that there is no strictly smaller interpretation $N \subset M_X$ with this property. This is an additional feature brought along the minimality of stable models. As a consequence, the DLP-function $\Pi_n^{\text{sat}}$ does not capture all possible truth assignments to variables in $X$ but no relevant truth assignments are lost. On the other hand, any stable model $M$ of $\Pi_n^{\text{unsat}}$ indicates that the respective theory

$$\{C_i \vee \neg D_i \mid 1 \leq i \leq n, \text{act}(i) \in M\}$$

is inconsistent, or alternatively, the formula $\bigvee_{1 \leq i \leq n, \text{act}(i) \in M} \neg C_i \wedge D_i$ is valid.

Concerning the correctness of the representation given in Figure 2, due to an existing proof by Janhunen et al. (2006), we only present the main steps—fully exploiting the benefits from our modular approach.

**Theorem 5.14**  *A QBF $\exists X \forall Y \phi$ of the form (19) is valid iff $\text{SM}(\Pi_n^{\text{sat}} \sqcup \Pi_n^{\text{unsat}})$ is non-empty.*

*Proof sketch.* Consider any QBF $\exists X \forall Y \phi$ of the form (19). The following are equivalent:

1. The formula $\exists X \forall Y \phi$ is valid.

2. There is a minimal interpretation $N \subseteq X$ such that, for the set $I = \{1 \leq i \leq n \mid N \not\models A_i \vee \neg B_i\}$ of indices determined by $N$ with $N \models \{\neg A_i \wedge B_i \mid i \in I\} \cup \{A_i \vee \neg B_i \mid i \notin I\}$, the theory $\{C_i \vee \neg D_i \mid i \in I\}$ is unsatisfiable.

3. The DLP-functions $\Pi_n^{\mathrm{sat}}$ and $\Pi_n^{\mathrm{unsat}}$ have compatible stable models $M_1 = N \cup \{\mathsf{act}(i) \mid i \in I\}$ and $M_2 = \{\mathsf{act}(i) \mid i \in I\} \cup Y \cup \{u\}$, respectively.

4. The DLP-function $\Pi_n^{\mathrm{sat}} \sqcup \Pi_n^{\mathrm{unsat}}$ has a stable model

$$M = M_1 \cup M_2 = N \cup \{\mathsf{act}(i) \mid i \in I\} \cup Y \cup \{u\}.$$

In the second item, the minimality of $N$ means that there is no $N' \subset N$ such that $\{1 \leq i \leq n \mid N' \not\models A_i \vee \neg B_i\} = I$. This can be assumed without loss of generality. □

Theorem 5.14 and the module theorem suggest an approximation strategy for verifying the validity of QBFs of the form (19). If either $\mathrm{SM}(\Pi_n^{\mathrm{sat}})$ or $\mathrm{SM}(\Pi_n^{\mathrm{unsat}})$ is empty, we know directly that the formula is not valid. Otherwise, we check whether $\mathrm{SM}(\Pi_n^{\mathrm{sat}} \sqcup \Pi_n^{\mathrm{unsat}}) = \emptyset$.

## 5.3 Splitting Sets

For the sake of comparison, we formulate the splitting-set theorem (Lifschitz & Turner, 1994) for a DLP-function $\Pi = \langle R, \emptyset, O, \emptyset \rangle$, which essentially forms an "ordinary" disjunctive program. Splitting sets are sets of atoms that are closed in the following sense:

**Definition 5.15** *Given a DLP-function $\Pi = \langle R, \emptyset, O, \emptyset \rangle$, a set $U \subseteq O$ of atoms is a splitting set for $\Pi$ if and only if, for every rule $A \leftarrow B, {\sim}C \in R$,*

$$A \cap U \neq \emptyset \text{ implies } A \cup B \cup C \subseteq U.$$

By Definitions 2.1 and 5.15, the sets $\emptyset$ and $O$ are always splitting sets for $\Pi$. However, one is mostly interested in other *non-trivial* splitting sets $\emptyset \subset U \subset O$ for $\Pi$, but such sets need not exist. Nevertheless, any splitting set $U$ divides the respective set of rules $R$ in two parts. The *bottom*, $\mathrm{b}_U(R)$, of $R$ with respect to $U$ contains all rules $A \leftarrow B, {\sim}C \in R$ such that $A \cup B \cup C \subseteq U$, whereas the *top*, $\mathrm{t}_U(R)$, of $R$ is $R \setminus \mathrm{b}_U(R)$. The splitting of $R$ into $\mathrm{b}_U(R)$ and $\mathrm{t}_U(R)$ becomes a proper one, i.e., $\mathrm{b}_U(R) \neq \emptyset$ and $\mathrm{t}_U(R) \neq \emptyset$, if

1. $U$ is non-trivial and

2. every atom $a \in O$ has at least one defining rule $A \leftarrow B, {\sim}C \in R$ such that $a \in A$.

According to Lifschitz and Turner (1994), a *solution* to $R$ with respect to $U \subseteq O$ is a pair $\langle X, Y \rangle$ where $X \subseteq U$, $Y \subseteq O \setminus U$, $X \in \mathrm{SM}(\mathrm{b}_U(R))$, and $Y \in \mathrm{SM}(\mathrm{t}_U(R)/X)$. Here, $\mathrm{t}_U(R)/X$ denotes the *partial evaluation* of $\mathrm{t}_U(R)$ in the sense of Definition 3.2 using $X \subseteq U$ as an input interpretation. Using a similar idea, let us introduce DLP-functions corresponding to $\mathrm{b}_U(R)$ and $\mathrm{t}_U(R)$. Given a splitting set $U$ for $\Pi$, the join $\Pi = \Pi_B \sqcup \Pi_T$, where

$$\Pi_B = \langle \mathrm{b}_U(R), \emptyset, U, \emptyset \rangle \text{ and } \Pi_T = \langle \mathrm{t}_U(R), U, O \setminus U, \emptyset \rangle$$

is defined. Then, the following result is implied by Theorem 5.7.

**Corollary 5.16 (Splitting-Set Theorem from Lifschitz & Turner, 1994)** *For every DLP-function $\Pi = \langle R, \emptyset, O, \emptyset \rangle$ corresponding to a set $R$ of disjunctive rules, every splitting set $U \subseteq O$ for $\Pi$, and every interpretation $M \subseteq \mathrm{At}(\Pi) = O$, the following conditions are equivalent:*

1. *$M$ is a stable model of $\Pi$.*

2. *$M \cap U \in \mathrm{SM}(\Pi_B)$ and $M \in \mathrm{SM}(\Pi_T)$.*

3. *$\langle M \cap U, M \setminus U \rangle$ is a solution to $R$ with respect to $U$.*

In fact, Theorem 5.7 is strictly stronger than the splitting-set theorem. As previously demonstrated by Oikarinen and Janhunen (2008a), splitting sets are applicable to DLP-functions like $\Pi = \langle \{ a \leftarrow \sim b; \; b \leftarrow \sim a \}, \emptyset, \{a, b\}, \emptyset \rangle$ only in the trivial way, i.e., only $U_1 = \emptyset$ are $U_2 = \{a, b\}$ are splitting sets for $\Pi$. In contrast, Theorem 5.7 applies to the preceding DLP-function in more versatile ways, i.e., $\Pi_1 \sqcup \Pi_2$ is defined for $\Pi_1 = \langle \{ a \leftarrow \sim b \}, \{b\}, \{a\}, \emptyset \rangle$ and $\Pi_2 = \langle \{ b \leftarrow \sim a \}, \{a\}, \{b\}, \emptyset \rangle$. As a consequence of $\Pi_1 \sqcup \Pi_2$ being defined, it is possible to determine the sets of stable models $\mathrm{SM}(\Pi_1) = \{\{a\}, \{b\}\} = \mathrm{SM}(\Pi_2)$ in separation, if appropriate, and then conclude that $\mathrm{SM}(\Pi) = \mathrm{SM}(\Pi_1) \bowtie \mathrm{SM}(\Pi_2) = \{\{a\}, \{b\}\}$ holds as well. Yet another generality aspect of splitting concerns the role of input atoms—they are assumed nonexistent above. Theorem 5.7, however, enables us to treat them as well.

## 6. Decomposing DLP-Functions

The objectives of this section are contrary to the construction of a DLP-function as a join of modules. The idea is to exploit the strongly connected components of $\mathrm{DG}^+(\Pi)$, for a DLP-function $\Pi$, in order to *decompose* $\Pi$ into smaller components, e.g., when there is no *a priori* information about the internal structure of $\Pi$. For simplicity, we will first consider DLP-functions $\Pi$ having no hidden atoms, i.e., where $\mathrm{At}_\mathrm{h}(\Pi) = \emptyset$. The effects of hidden atoms on the decomposition of DLP-functions will be addressed thereafter. As defined in conjunction with Definition 2.6, the SCCs in $\mathrm{DG}^+(\Pi)$ are induced by the positive dependency relation $\leq$ which is reflexive and transitive, i.e., a *preorder* by definition. In the sequel, the set of SCCs in $\mathrm{DG}^+(\Pi)$ is denoted by $\mathrm{SCC}^+(\Pi)$. The positive dependency relation $\leq$ lifts for the elements of $\mathrm{SCC}^+(\Pi)$ as follows: $S_1 \leq S_2$ if and only if there are atoms $a_1 \in S_1$ and $a_2 \in S_2$ such that $a_1 \leq a_2$. To this end, it does not matter which pair of atoms is inspected.

**Lemma 6.1** *For any DLP-function $\Pi$ and any components $S_1, S_2 \in \mathrm{SCC}^+(\Pi)$, $S_1 \leq S_2$ if and only if $a_1 \leq a_2$ for every $a_1 \in S_1$ and $a_2 \in S_2$.*

*Proof.* ($\Longrightarrow$) If $S_1 \leq S_2$, there are $b_1 \in S_1$ and $b_2 \in S_2$ such that $b_1 \leq b_2$. Consider any $a_1 \in S_1$ and $a_2 \in S_2$. It follows that $a_1 \leq b_1$ and $b_2 \leq a_2$ by the definition of SCCs. Thus, $a_1 \leq a_2$ as $\leq$ is transitive.

($\Longleftarrow$) This holds trivially as SCCs are non-empty. $\square$

**Proposition 6.2** *The relation $\leq$ over $\mathrm{SCC}^+(\Pi)$ is reflexive, transitive, and antisymmetric.*

*Proof.* The relation $\leq$ over $\mathrm{SCC}^+(\Pi)$ is reflexive and transitive by definition. For antisymmetry, consider any $S_1, S_2 \in \mathrm{SCC}^+(\Pi)$ such that $S_1 \leq S_2$ and $S_2 \leq S_1$. It follows by Lemma 6.1 that, for every $a_1 \in S_1$ and $a_2 \in S_2$, $a_1 \leq a_2$ and $a_2 \leq a_1$. Thus, $S_1 = S_2$ by the maximality of components in $\mathrm{SCC}^+(\Pi)$. □

Consequently, we may conclude that $\langle \mathrm{SCC}^+(\Pi), \leq \rangle$ is a *partially ordered set*. Since $\Pi$ is finite by definition, $\langle \mathrm{SCC}^+(\Pi), \leq \rangle$ has maxima and minima but these elements need not be unique. In particular, for each $S \in \mathrm{SCC}^+(\Pi)$ there is a minimum element $S_1 \in \mathrm{SCC}^+(\Pi)$ such that $S_1 \leq S$ and $S_2 \leq S_1$ implies $S_2 = S_1$, for any $S_2 \in \mathrm{SCC}^+(\Pi)$. Thus, we may apply the principle of *well-founded induction* using the minima of $\langle \mathrm{SCC}^+(\Pi), \leq \rangle$ as basis.

Given the structure $\langle \mathrm{SCC}^+(\Pi), \leq \rangle$, the DLP-function $\Pi = \langle R, I, O, \emptyset \rangle$ can be decomposed in the following way: The set of rules associated with $S \in \mathrm{SCC}^+(\Pi)$ is $\mathrm{Def}_R(S)$ from (2), i.e., the set of *defining rules* for $S$ in $R$. In general, the head of an arbitrary rule $A \leftarrow B, \sim C \in R$ may coincide in the sense of (2) with several SCCs, which implies that the rule is included in $\mathrm{Def}_R(S)$ for several $S \in \mathrm{SCC}^+(\Pi)$. However, such a distribution of rules is in perfect harmony with the last two conditions of Definition 2.2. We must also bear in mind integrity constraints $\perp \leftarrow B, \sim C$ which are not included in $\mathrm{Def}_R(S)$ for any $S \in \mathrm{SCC}^+(\Pi)$. To access the integrity constraints of any set $R$ of rules, we define

$$\mathrm{IC}(R) = \{A \leftarrow B, \sim C \in R \mid A = \emptyset\}. \tag{23}$$

We are now ready to present a decomposition of $\Pi$ based on $\mathrm{SCC}^+(\Pi)$.

**Definition 6.3** *Given a DLP-function* $\Pi = \langle R, I, O, \emptyset \rangle$, *the* decomposition induced by $\mathrm{SCC}^+(\Pi)$ *includes a DLP-function*

$$\Pi_0 = \langle \mathrm{IC}(R), \mathrm{At}(\mathrm{IC}(R)) \cup (I \setminus \mathrm{At}(R)), \emptyset, \emptyset \rangle \tag{24}$$

*and, for each* $S \in \mathrm{SCC}^+(\Pi)$, *a DLP-function*

$$\Pi_S = \langle \mathrm{Def}_R(S), \mathrm{At}(\mathrm{Def}_R(S)) \setminus S, S, \emptyset \rangle. \tag{25}$$

The purpose of the extra module $\Pi_0$ is to keep track of integrity constraints as well as input atoms that are not mentioned by the rules of $R$. The other modules involved in the decomposition $\Pi$ are induced by SCCs. Each $\Pi_S$ refers to other modules using $\mathrm{At}(\mathrm{Def}_R(S)) \setminus S$ as its input signature and provides the defining rules (if any) for every atom in $S$. Recall that an output atom having no defining rules will be falsified by default.

**Proposition 6.4** *For a DLP-function* $\Pi = \langle R, I, O, \emptyset \rangle$ *and its decomposition based on* $\mathrm{SCC}^+(\Pi)$, *the join*

$$\Pi_0 \sqcup (\bigsqcup_{S \in \mathrm{SCC}^+(\Pi)} \Pi_S) \tag{26}$$

*is defined and equal to* $\Pi$.

*Proof.* Let us consider $\Pi_0$ and $\Pi_S$ for any $S \in \mathrm{SCC}^+(\Pi)$. The composition $\Pi_0 \oplus \Pi_S$ is defined because these modules involve no hidden atoms, $\mathrm{At}_o(\Pi_0) = \emptyset$, and we have $\mathrm{Def}_{R_1}(\emptyset) = \emptyset = \mathrm{Def}_{R_1 \cup R_2}(\emptyset)$ and $\mathrm{Def}_{R_2}(S) = \mathrm{Def}_R(S) = \mathrm{Def}_{R_1 \cup R_2}(S)$ for the sets of rules $R_1 = \mathrm{IC}(R)$ and $R_2 = \mathrm{Def}_R(S)$. The join $\Pi_0 \sqcup \Pi_S$ is defined as the respective composition is and the integrity constraints in $\Pi_0$ do not create any dependencies in $\mathrm{DG}^+(\Pi_0 \oplus \Pi_S)$.

Let us perform a similar analysis for $\Pi_{S_1}$ and $\Pi_{S_2}$ based on two different components $S_1, S_2 \in \mathrm{SCC}^+(\Pi)$. It is clear that $\Pi_{S_1} \oplus \Pi_{S_2}$ is defined since these modules involve no hidden atoms, $S_1 \cap S_2 = \emptyset$, and we have that $\mathrm{Def}_{R_1}(S_1) = \mathrm{Def}_R(S_1) = \mathrm{Def}_{R_1 \cup R_2}(S_1)$ and $\mathrm{Def}_{R_2}(S_2) = \mathrm{Def}_R(S_2) = \mathrm{Def}_{R_1 \cup R_2}(S_2)$, for $R_1 = \mathrm{Def}_R(S_1)$ and $R_2 = \mathrm{Def}_R(S_2)$.

Since all pairwise joins are defined, also the overall join (26) is defined. By Definition 2.4 and the definition of $\mathrm{SCC}^+(\Pi)$, the outcome is equal to $\Pi$ because

1. $\mathrm{IC}(R) \cup \bigcup_{S \in \mathrm{SCC}^+(\Pi)} \mathrm{Def}_R(S) = R$,

2. $\bigcup_{S \in \mathrm{SCC}^+(\Pi)} S = O$, and

3. $(\mathrm{At}(\mathrm{IC}(R)) \setminus O) \cup ((I \setminus \mathrm{At}(R)) \setminus O) \cup \bigcup_{S \in \mathrm{SCC}^+(\Pi)} (\mathrm{At}(\mathrm{Def}_R(S)) \setminus O) = I$.     $\square$

**Corollary 6.5** *For a DLP-function $\Pi$ with $\mathrm{At}_\mathrm{h}(\Pi) = \emptyset$ and its decomposition based on $\mathrm{SCC}^+(\Pi)$,*

$$\mathrm{SM}(\Pi) = \mathrm{SM}(\Pi_0) \bowtie (\bigboxtimes_{S \in \mathrm{SCC}^+(\Pi)} \mathrm{SM}(\Pi_S)).$$

**Example 6.6** *Consider the following DLP-function $\Pi$:*

| $\{a, b, c, d\}$ | |
| :--- | :--- |
| $a \vee b \vee c \vee d \leftarrow;$ | |
| $\leftarrow a, c;$ | $\leftarrow b, c;$ |
| $\leftarrow a, d;$ | $\leftarrow b, d;$ |
| $a \leftarrow b;$ | $c \leftarrow d;$ |
| $b \leftarrow a;$ | $d \leftarrow c.$ |
| $\emptyset$ | |

*So, $\mathrm{At}_\mathrm{i}(\Pi) = \emptyset$, $\mathrm{At}_\mathrm{o}(\Pi) = \{a, b, c, d\}$, and $\mathrm{At}_\mathrm{h}(\Pi) = \emptyset$. There are two SCCs in $\mathrm{DG}^+(\Pi)$, viz. $S_1 = \{a, b\}$ and $S_2 = \{c, d\}$. The resulting decomposition of $\Pi$ consists of*

$$\Pi_0 = \langle \{\leftarrow a, c; \leftarrow a, d; \leftarrow b, c; \leftarrow b, d\}, \{a, b, c, d\}, \emptyset, \emptyset \rangle,$$
$$\Pi_{S_1} = \langle \{a \vee b \vee c \vee d \leftarrow; a \leftarrow b; b \leftarrow a\}, \{c, d\}, \{a, b\}, \emptyset \rangle, \text{ and}$$
$$\Pi_{S_2} = \langle \{a \vee b \vee c \vee d \leftarrow; c \leftarrow d; d \leftarrow c\}, \{a, b\}, \{c, d\}, \emptyset \rangle.$$

*The respective sets of stable models are*

$$\mathrm{SM}(\Pi_0) = \{\{a, b\}, \{c, d\}, \{a\}, \{b\}, \{c\}, \{d\}, \emptyset\},$$
$$\mathrm{SM}(\Pi_{S_1}) = \{\{a, b\}, \{c\}, \{d\}, \{c, d\}\},$$
$$\mathrm{SM}(\Pi_{S_1}) = \{\{c, d\}, \{a\}, \{b\}, \{a, b\}\}, \text{ and}$$
$$\mathrm{SM}(\Pi) = \{\{a, b\}, \{c, d\}\}.$$

Next, we address the case of DLP-functions involving hidden atoms, i.e., for which $\mathrm{At}_\mathrm{h}(\Pi) \neq \emptyset$ holds. Then, the components in $\mathrm{DG}^+(\Pi)$ are subsets of $\mathrm{At}_\mathrm{o}(\Pi) \cup \mathrm{At}_\mathrm{h}(\Pi)$ and we have to revise (25) accordingly. For a DLP-function $\Pi = \langle R, I, O, H \rangle$ and $S \in \mathrm{SCC}^+(\Pi)$,

$$\Pi_S = \langle \mathrm{Def}_R(S), \mathrm{At}(\mathrm{Def}_R(S)) \setminus S, S \cap O, S \cap H \rangle. \tag{27}$$

Unfortunately, a decomposition based on modules of the form (27) is likely to be too fine-grained. For certain components $S_1, S_2 \in \mathrm{SCC}^+(\Pi)$ such that $S_1 \neq S_2$, the respective modules $\Pi_{S_1}$ and $\Pi_{S_2}$ conforming to (27) might not respect hidden atoms of each other. A similar setting may arise with $\Pi_0$ and an individual module $\Pi_S$ based on some $S \in \text{SCC}^+(\Pi)$ if the integrity constraints of $\Pi$ refer to hidden atoms of $\Pi_S$. The problem would disappear if all hidden atoms of $\Pi$ were revealed but this is hardly appropriate—there are good reasons to hide certain atoms from a knowledge representation perspective.

A way to approach this problem is to distinguish components $S_1 \in \text{SCC}^+(\Pi)$ and $S_2 \in \text{SCC}^+(\Pi)$ for which the respective modules $\Pi_{S_1}$ and $\Pi_{S_2}$ would not respect the hidden atoms of each other, i.e., a hidden atom defined by one would be referred by the other—either positively or negatively. Similar conflicts could also arise due to integrity constraints packed into the module $\Pi_0$ distinguished in Definition 6.3. At first sight, we should amalgamate $\Pi_0$ with any other module whose hidden atoms occur in the integrity constraints of $\Pi_0$. But, in order to avoid fusions of this kind as far as possible, it is worth redistributing integrity constraints referring to hidden atoms. This is clearly possible for integrity constraints referring to hidden atoms involved in a single component only. To formalize the ideas presented so far, we distinguish a precise relation among the components of $\text{SCC}^+(\Pi)$ as follows.

**Definition 6.7** *Given a DLP-function $\Pi$, components $S_1, S_2 \in \text{SCC}^+(\Pi)$ do not respect the hidden atoms of each other, denoted by $S_1 \leftrightsquigarrow_h S_2$, if and only if $S_1 \neq S_2$ and there is*

1. *a hidden atom $h \in \text{At}_h(\Pi_{S_1})$ such that $h \in \text{At}_i(\Pi_{S_2})$, or*

2. *a hidden atom $h \in \text{At}_h(\Pi_{S_2})$ such that $h \in \text{At}_i(\Pi_{S_1})$, or*

3. *there are hidden atoms $h_1 \in \text{At}_h(\Pi_{S_1})$ and $h_2 \in \text{At}_h(\Pi_{S_2})$ which both have an occurrence in some integrity constraint $\leftarrow B, \sim C$ of $\Pi$.*

It is clear that $\leftrightsquigarrow_h$ is irreflexive and symmetric for the components of $\text{SCC}^+(\Pi)$ for any DLP-function $\Pi$. Moreover, the transitive closure of $\leftrightsquigarrow_h$, denoted by $\leftrightsquigarrow_h^+$, gives rise to a repartitioning of $\text{SCC}^+(\Pi)$. Each maximal block $S_1, \ldots, S_n$ of components such that $S_i \leftrightsquigarrow_h^+ S_j$ holds for every $i \neq j$ induces a module $\Pi_S$ as determined by (27) for the union $S = S_1 \cup \ldots \cup S_n$. The key observation is that modules associated with different blocks of components respect hidden atoms of each other which makes Theorem 5.7 applicable at that level of abstraction. To summarize the treatment of DLP-functions involving hidden atoms in their rules, we revise Definition 6.3 accordingly.

**Definition 6.8** *Given a DLP-function $\Pi = \langle R, I, O, H \rangle$, the decomposition induced by $\text{SCC}^+(\Pi)$ and $\leftrightsquigarrow_h^+$ includes a DLP-function*

$$\Pi_0 = \langle \text{IC}_0(R), \text{At}(\text{IC}_0(R)) \cup (I \setminus \text{At}(R)), \emptyset, \emptyset \rangle \tag{28}$$

*where $\text{IC}_0(R) = \{\leftarrow B, \sim C \in R \mid (B \cup C) \cap H = \emptyset\}$ and, for each maximal block $S_1, \ldots, S_n$ of components of $\text{SCC}^+(\Pi)$ such that $S_i \leftrightsquigarrow_h^+ S_j$ for every $i \neq j$, a DLP-function*

$$\Pi_S = \langle \text{Def}_R(S) \cup \text{IC}_S(R), \text{At}(\text{Def}_R(S) \cup \text{IC}_S(R)) \setminus S, S \cap O, S \cap H \rangle \tag{29}$$

*where $S = S_1 \cup \ldots \cup S_n$ and $\text{IC}_S(R) = \{\leftarrow B, \sim C \in R \mid (B \cup C) \cap (S \cap H) \neq \emptyset\}$.*

As regards Example 6.6, Definitions 6.3 and 6.8 yield identical decompositions for the DLP-function in question. The effects of hiding are demonstrated by the following example:

**Example 6.9** *Consider a DLP-function $\Pi = \langle R, \emptyset, O, H \rangle$, for*

$$R = \{\leftarrow a, \sim c; \ a \vee b \leftarrow; \ b \vee c \vee d \leftarrow; \ c \leftarrow d; \ d \leftarrow c, \sim b\}$$

*and $O \cup H = \{a, b, c, d\}$, where the exact partitioning of atoms in $O$ and $H$ varies from case to case as analyzed below. The SCCs in $\mathrm{SCC}^+(\Pi)$ are $S_1 = \{a\}$, $S_2 = \{b\}$, and $S_3 = \{c, d\}$.*

1. *If we take all atoms visible in $\Pi$, i.e., if $H = \emptyset$, the decomposition of $\Pi$ yields three modules, $\Pi_{S_1} = \langle \{a \vee b \leftarrow\}, \{b\}, \{a\}, \emptyset \rangle$, $\Pi_{S_2} = \langle \{a \vee b \leftarrow; \ b \vee c \vee d \leftarrow\}, \{a, c, d\}, \{b\}, \emptyset \rangle$, and $\Pi_{S_3} = \langle \{b \vee c \vee d \leftarrow; \ c \leftarrow d; \ d \leftarrow c, \sim b\}, \{b\}, \{c, d\}, \emptyset \rangle$, in addition to the module $\Pi_0 = \langle \{\leftarrow a, \sim c\}, \{a, c\}, \emptyset, \emptyset \rangle$ encompassing integrity constraints.*

2. *If we hide $H = \{a\}$ in $\Pi$, we obtain $S_1 \leftrightsquigarrow_{\mathrm{h}} S_2$ by the disjunctive rule $a \vee b \leftarrow$. Therefore, components $S_1$ and $S_2$ must be placed in the same block which is also maximal—giving rise to a module $\Pi_S = \langle \{\leftarrow a, \sim c; \ a \vee b \leftarrow; \ b \vee c \vee d \leftarrow\}, \{c, d\}, \{b\}, \{a\} \rangle$ where $S = S_1 \cup S_2 = \{a, b\}$. The other modules are $\Pi_0 = \varnothing$ and $\Pi_{S_3}$ listed above.*

3. *Finally, if we set $H = \{a, c\}$ for $\Pi$, we obtain $S_2 \leftrightsquigarrow_{\mathrm{h}} S_3$ by $b \vee c \vee d \leftarrow$ and $S_1 \leftrightsquigarrow_{\mathrm{h}} S_3$ by $\leftarrow a, \sim c$ in addition to $S_1 \leftrightsquigarrow_{\mathrm{h}} S_2$ as stated above. Since $\Pi_0 = \varnothing$, the decomposition of $\Pi$ effectively collapses to a single module $\Pi_{S'} = \Pi$ where $S' = S_1 \cup S_2 \cup S_3$.*

*We note about the non-trivial modules mentioned above that*

$$\begin{aligned}
\mathrm{SM}(\Pi_{S_1}) &= \{\{a\}, \{b\}\}, \\
\mathrm{SM}(\Pi_{S_2}) &= \{\{b\}, \{a, b\}, \{b, c\}, \{a, c\}, \{b, d\}, \{a, d\}, \{b, c, d\}, \{a, c, d\}\}, \\
\mathrm{SM}(\Pi_{S_3}) &= \{\{b\}, \{c, d\}\}, \\
\mathrm{SM}(\Pi_0) &= \{\emptyset, \{c\}, \{a, c\}\}, \ \text{and} \\
\mathrm{SM}(\Pi_S) &= \{\{b\}, \{a, c\}, \{b, c\}, \{b, d\}, \{a, c, d\}, \{b, c, d\}\}.
\end{aligned}$$

*But, regardless of the decomposition obtained, it holds for the respective joins that*

$$\begin{aligned}
\mathrm{SM}(\Pi) &= \mathrm{SM}(\Pi_{S_1}) \bowtie \mathrm{SM}(\Pi_{S_2}) \bowtie \mathrm{SM}(\Pi_{S_3}) \bowtie \mathrm{SM}(\Pi_0) \\
&= \mathrm{SM}(\Pi_S) \bowtie \mathrm{SM}(\Pi_{S_3}) \bowtie \mathrm{SM}(\varnothing) \\
&= \mathrm{SM}(\Pi_{S'}) \bowtie \mathrm{SM}(\varnothing) \\
&= \{\{a, c, d\}, \{b\}\}.
\end{aligned}$$

In the calculations involving $\bowtie$ it is important to notice that the allowed combinations of stable models are determined in terms of *joint visible* atoms of the modules involved. For instance, we have $\mathrm{At_v}(\Pi_{S_1}) \cap \mathrm{At_v}(\Pi_{S_3}) = \{a, b\} \cap \{b, c, d\} = \{b\}$ so that $\mathrm{SM}(\Pi_{S_1}) \bowtie \mathrm{SM}(\Pi_{S_3})$ is $\{\{a\} \cup \{c, d\}, \{b\} \cup \{b\}\} = \{\{a, c, d\}, \{b\}\}$ by Definition 5.2. Thus, interestingly, the role of the remaining two modules $\Pi_{S_2}$ and $\Pi_0$ is merely to approve upon these two models. Recalling the discussion from the introduction, this suggests a strategy which gives precedence to

1. an evaluation of modules having only few stable models, and

2. a combination of stable models for modules that have only few visible atoms in common.

## 7. Shifting Disjunctions

In this section, we continue the pursuit of applications for the module theorem established in Section 5. We now generalize the principle of *shifting* disjunctive rules (Gelfond et al., 1991; Dix et al., 1996) by applying the results of this paper. Roughly speaking, the idea behind shifting is to translate a disjunctive rule $A \leftarrow B, \sim C$ into several normal (non-disjunctive) rules by shifting head atoms $h \in A$ to negative literals $\sim h$ in the body. For instance, a simple disjunctive rule $a \lor b \lor c \leftarrow$ is captured by normal rules

$$a \leftarrow \sim b, \sim c, \quad b \leftarrow \sim a, \sim c, \quad \text{and} \quad c \leftarrow \sim a, \sim b.$$

As shown by Eiter et al. (2004), such a local shifting transformation preserves *ordinary equivalence*, i.e., stable models.[12] The application of this technique is, however, pre-empted in the presence of *head-cycles* (Ben-Eliyahu & Dechter, 1994). Such a cycle is provided by an SCC $S$ that intersects with the head $A$ of some disjunctive rule $A \leftarrow B, \sim C$ of $\Pi$ such that $|S \cap A| > 1$. For instance, local shifting is no longer applicable to the rule $a \lor b \lor c \leftarrow$ in the presence of $a \leftarrow b$ and $b \leftarrow a$ which create a strongly connected component $S = \{a, b\}$. As a consequence, the respective DLP-functions

$$\Pi_1 = \langle \{a \lor b \lor c \leftarrow; \ a \leftarrow b; \ b \leftarrow a\}, \emptyset, \{a, b, c\}, \emptyset \rangle, \tag{30}$$

$$\Pi_2 = \langle \{a \leftarrow \sim b, \sim c; \ b \leftarrow \sim a, \sim c; \ c \leftarrow \sim a, \sim b; \ a \leftarrow b; \ b \leftarrow a\}, \emptyset, \{a, b, c\}, \emptyset \rangle \tag{31}$$

have different stable models: $\mathrm{SM}(\Pi_1) = \{\{a, b\}, \{c\}\}$ and $\mathrm{SM}(\Pi_2) = \{\{c\}\}$. Such a discrepancy of stable models can be settled by applying the decomposition technique from Section 6. In fact, it leads to a proper generalization of the local shifting transformation which is formalized below for DLP-functions and their strongly connected components.

**Definition 7.1** *Let $\Pi = \langle R, I, O, H \rangle$ be a DLP-function and $\mathrm{SCC}^+(\Pi)$ the respective set of SCCs. The* general shifting *of $\Pi$ is the DLP-function $\mathrm{GSH}(\Pi) = \langle \mathrm{IC}(R) \cup R', I, O, H \rangle$, where $R'$ is the set of rules*

$$\{(A \cap S) \leftarrow B, \sim C, \sim (A \setminus S) \mid A \leftarrow B, \sim C \in R, \ S \in \mathrm{SCC}^+(\Pi) \ and \ A \cap S \neq \emptyset\}. \tag{32}$$

Hence, the idea is to project the head $A$ of the rule with respect to each component $S$, and atoms in the difference $A \setminus S$ are shifted to the negative body. This can be viewed as the contribution of a disjunctive rule $A \leftarrow B, \sim C$ for a particular component $S$.

**Example 7.2** *For $\Pi_1$ from (30), we have $\mathrm{SCC}^+(\Pi_1) = \{\{a, b\}, \{c\}\}$, so that*

$$\mathrm{GSH}(\Pi_1) = \langle \{a \lor b \leftarrow \sim c; \ c \leftarrow \sim a, \sim b; \ a \leftarrow b; \ b \leftarrow a\}, \emptyset, \{a, b, c\}, \emptyset \rangle.$$

*Most importantly, we have $\mathrm{SM}(\mathrm{GSH}(\Pi_1)) = \{\{a, b\}, \{c\}\} = \mathrm{SM}(\Pi_1)$, in contrast to the set $\mathrm{SM}(\Pi_2) = \{\{c\}\}$ of stable models for $\Pi_2$ from (31).*

---

12. In addition to ordinary equivalence, also *uniform equivalence* (Eiter & Fink, 2003) is preserved by local shifting but not *strong equivalence* (Lifschitz, Pearce, & Valverde, 2001).

We now prove the correctness of the general shifting principle from Definition 7.1. The aim is to exploit the decomposition of $\Pi$ from Definition 6.3 together with the modular reconstruction of $\Pi$ from Proposition 6.4 and the compositionality of stable semantics from Corollary 6.5. To extend the coverage of Corollary 6.5, we introduce explicit operators for *revealing* and *hiding* atoms of DLP-functions as follows:

**Definition 7.3** *Let* $\Pi = \langle R, I, O, H \rangle$ *be a DLP-function. Then,*

1. *Reveal*$(\Pi, A) = \langle R, I, O \cup A, H \setminus A \rangle$, *for a set* $A \subseteq H$ *of hidden atoms, and*

2. *Hide*$(\Pi, A) = \langle R, I, O \setminus A, H \cup A \rangle$, *for a set* $A \subseteq O$ *of output atoms.*

Since the definition of stable models does not make a difference between output atoms and hidden atoms, the following properties are easy to verify. The role of hidden atoms becomes important in Section 8 when DLP-functions are compared with each other.

**Proposition 7.4** *Let* $\Pi$ *be any DLP-function.*

1. *For any* $A \subseteq \mathrm{At_h}(\Pi)$, $\mathrm{SM}(\Pi) = \mathrm{SM}(\mathrm{Reveal}(\Pi, A))$.

2. *For any* $A \subseteq \mathrm{At_o}(\Pi)$, $\mathrm{SM}(\Pi) = \mathrm{SM}(\mathrm{Hide}(\Pi, A))$.

**Lemma 7.5** *Let* $\Pi$ *be a DLP-function with* $\mathrm{At_h}(\Pi) = \emptyset$, $S$ *a component in* $\mathrm{SCC}^+(\Pi)$, *and* $\Pi_S$ *the respective module in the decomposition of* $\Pi$ *according to Definition 6.3. Then,*

$$\mathrm{SM}(\Pi_S) = \mathrm{SM}(\mathrm{GSH}(\Pi_S)). \tag{33}$$

*Proof.* Recall that $\Pi_S = \langle \mathrm{Def}_R(S), I, S, \emptyset \rangle$, where the input signature $I = \mathrm{At}(\mathrm{Def}_R(S)) \setminus S$. Notice that $S$ is the only component in $\mathrm{SCC}^+(\Pi_S)$ and hence $\mathrm{GSH}(\Pi_S)$ has a set of rules

$$R' = \{(A \cap S) \leftarrow B, \sim C, \sim(A \setminus S) \mid A \leftarrow B, \sim C \in \mathrm{Def}_R(S)\}.$$

Consider any interpretation $M \subseteq I \cup S$, where $I$ and $S$ are the input and output signatures of $\Pi_S$, respectively. Thus, $M_\mathrm{i} = M \cap I$ and $M_\mathrm{o} = M \cap S$. Then, the following equivalences hold:

$$
\begin{aligned}
& A \leftarrow B \in (\mathrm{Def}_R(S)/M_\mathrm{i})^{M_\mathrm{o}} \\
\Longleftrightarrow\ & \exists A \leftarrow B, \sim C \in \mathrm{Def}_R(S)/M_\mathrm{i} \text{ such that } M_\mathrm{o} \models \sim C \\
\Longleftrightarrow\ & \exists A' \leftarrow B', \sim C' \in \mathrm{Def}_R(S) \text{ such that } A = A'_\mathrm{o},\ B = B'_\mathrm{o},\ C = C'_\mathrm{o}, \\
& M_\mathrm{i} \models \sim A'_\mathrm{i} \cup B'_\mathrm{i} \cup \sim C'_\mathrm{i}, \text{ and } M_\mathrm{o} \models \sim C'_\mathrm{o} \\
\Longleftrightarrow\ & \exists A \leftarrow B', \sim C', \sim A'_\mathrm{i} \in R' \text{ such that } A = A'_\mathrm{o},\ B = B'_\mathrm{o},\ C = C'_\mathrm{o}, \\
& M_\mathrm{i} \models \sim A'_\mathrm{i} \cup B'_\mathrm{i} \cup \sim C'_\mathrm{i}, \text{ and } M_\mathrm{o} \models \sim C'_\mathrm{o} \\
\Longleftrightarrow\ & \exists A \leftarrow B, \sim C \in R'/M_\mathrm{i} \text{ such that } M_\mathrm{o} \models \sim C \\
\Longleftrightarrow\ & A \leftarrow B \in (R'/M_\mathrm{i})^{M_\mathrm{o}}.
\end{aligned}
$$

Thus, we conclude that $(\mathrm{Def}_R(S)/M_\mathrm{i})^{M_\mathrm{o}}$ coincides with $(R'/M_\mathrm{i})^{M_\mathrm{o}}$, and, consequently, $M_\mathrm{o} \in \mathrm{MM}((\mathrm{Def}_R(S)/M_\mathrm{i})^{M_\mathrm{o}})$ if and only if $M_\mathrm{o} \in \mathrm{MM}((R'/M_\mathrm{i})^{M_\mathrm{o}})$. Therefore, $\mathrm{SM}(\Pi_S/M_\mathrm{i}) = \mathrm{SM}(\mathrm{GSH}(\Pi_S)/M_\mathrm{i})$. Since $M$ and, in particular, $M_\mathrm{i}$ were arbitrarily chosen in the beginning, we obtain the equality of stable models stated in (33) directly by Corollary 3.9. $\qquad\square$

**Theorem 7.6** *For any DLP-function* $\Pi = \langle R, I, O, H \rangle$, $\mathrm{SM}(\Pi) = \mathrm{SM}(\mathrm{GSH}(\Pi))$.

*Proof.* Since $\Pi$ may have hidden atoms, Corollary 6.5 is not applicable to its decomposition based on $\mathrm{SCC}^+(\Pi)$. Thus, we have to start with $\Pi' = \mathrm{Reveal}(\Pi, H) = \langle R, I, O \cup H, \emptyset \rangle$ rather than $\Pi$ itself. Since SCCs are independent of hiding, we have $\mathrm{SCC}^+(\Pi') = \mathrm{SCC}^+(\Pi)$ and $\mathrm{GSH}(\Pi') = \mathrm{Reveal}(\mathrm{GSH}(\Pi), H)$. Since $\mathrm{At_h}(\Pi') = \emptyset$ by construction, we know by Proposition 6.4 that $\Pi'_0 \sqcup (\bigsqcup_{S \in \mathrm{SCC}^+(\Pi)} \Pi'_S) = \Pi'$. Applying $\mathrm{GSH}(\cdot)$ to this equation yields

$$\mathrm{GSH}(\Pi') = \Pi'_0 \sqcup (\bigsqcup_{S \in \mathrm{SCC}^+(\Pi)} \mathrm{GSH}(\Pi'_S)). \tag{34}$$

As regards the respective sets of stable models, we obtain

$$
\begin{aligned}
\mathrm{SM}(\Pi') &= \mathrm{SM}(\Pi'_0) \bowtie (\bigbowtie_{S \in \mathrm{SCC}^+(\Pi')} \mathrm{SM}(\Pi'_S)) && \text{[Corollary 6.5]} \\
&= \mathrm{SM}(\Pi'_0) \bowtie (\bigbowtie_{S \in \mathrm{SCC}^+(\Pi')} \mathrm{SM}(\mathrm{GSH}(\Pi'_S))) && \text{[Lemma 7.5]} \\
&= \mathrm{SM}(\mathrm{GSH}(\Pi')). && \text{[Corollary 6.5 and (34)]}
\end{aligned}
$$

It follows by Proposition 7.4 that $\mathrm{SM}(\mathrm{Hide}(\Pi', H)) = \mathrm{SM}(\Pi') = \mathrm{SM}(\mathrm{GSH}(\Pi')) = \mathrm{SM}(\mathrm{Hide}(\mathrm{GSH}(\Pi'), H))$. Since $\mathrm{Hide}(\Pi', H) = \Pi$ and $\mathrm{Hide}(\mathrm{GSH}(\Pi'), H) = \mathrm{GSH}(\Pi)$, we have established that $\mathrm{SM}(\Pi) = \mathrm{SM}(\mathrm{GSH}(\Pi))$ as desired. ☐

According to Definition 6.3, decompositions of DLP-functions create multiple copies of disjunctive rules whose heads intersect with several SCCs. The introduction of such copies can be circumvented by applying the general shifting technique from Definition 7.1.

**Example 7.7** *For the DLP-function* $\Pi$ *from Example 6.6, we obtain* $R_1 = \{a \vee b \leftarrow \sim c, \sim d;\ a \leftarrow b;\ b \leftarrow a\}$ *and* $R_2 = \{c \vee d \leftarrow \sim a, \sim b;\ c \leftarrow d;\ d \leftarrow c\}$ *as the sets of rules associated with* $\Pi_1 = \langle R_1, \{c, d\}, \{a, b\}, \emptyset \rangle$ *and* $\Pi_2 = \langle R_2, \{a, b\}, \{c, d\}, \emptyset \rangle$, *for which* $\Pi_1 \sqcup \Pi_2 = \langle R_1 \cup R_2, \emptyset, \{a, b, c, d\}, \emptyset \rangle$ *is defined.* ☐

These observations enable us to view disjunctive rules which are shared by the modules associated with SCCs as syntactic sugar. However, a clever implementation can save space using shared rules. In the worst case, unwinding a rule $a_1 \vee \cdots \vee a_n \leftarrow B, \sim C$ that coincides with the respective SCCs $S_1, \ldots, S_n$ such that $a_1 \in S_1, \ldots, a_n \in S_n$ may create $n$ copies of the body $B \cup \sim C$. Such a quadratic blow-up can be partly alleviated by introducing a new atom $b$ as a name for the body. Thus the result of shifting $a_1 \in S_1, \ldots, a_n \in S_n$ becomes

$$
\begin{aligned}
&a_1 \leftarrow b, \sim a_2, \ldots, \sim a_n; \\
&\quad\quad \vdots \\
&a_i \leftarrow b, \sim a_1, \ldots, \sim a_{i-1}, \sim a_{i+1}, \ldots, \sim a_n; \\
&\quad\quad \vdots \\
&a_n \leftarrow b, \sim a_1, \ldots, \sim a_{n-1}
\end{aligned}
$$

together with the defining rule $b \leftarrow B, \sim C$ for $b$. There is an implementation of the general shifting principle called DENCODE.[13] If requested to do so, it calculates beforehand whether it pays off to introduce a new atom for the body for each disjunctive rule or not.

---

13. Available at `http://www.tcs.hut.fi/Software/asptools/` for experimenting.

## 8. Equivalence of DLP-Functions

The concept of *visible equivalence* was originally introduced in order to neglect hidden atoms when logic programs, or other theories of interest, are compared on the basis of their models (Janhunen, 2006). Oikarinen and Janhunen (2008a) extended this idea to the level of logic program modules—giving rise to the notion of *modular equivalence* for logic programs. In this section, we generalize the concept of modular equivalence for DLP-functions and introduce a translation-based method for checking modular equivalence of DLP-functions following analogous approaches of Oikarinen and Janhunen (2004, 2009).

### 8.1 Modular Equivalence

Module interfaces must be taken properly into account when DLP-functions are compared. For this reason, we consider two DLP-functions $\Pi_1$ and $\Pi_2$ to be *compatible* if and only if $\mathrm{At}_i(\Pi_1) = \mathrm{At}_i(\Pi_2)$ and $\mathrm{At}_o(\Pi_1) = \mathrm{At}_o(\Pi_2)$.

**Definition 8.1** *DLP-functions $\Pi_1$ and $\Pi_2$ are modularly equivalent, denoted by $\Pi_1 \equiv_m \Pi_2$, if and only if*

1. *$\Pi_1$ and $\Pi_2$ are compatible and*

2. *there is a bijection $f : \mathrm{SM}(\Pi_1) \to \mathrm{SM}(\Pi_2)$ such that for all interpretations $M \in \mathrm{SM}(\Pi_1)$, $M \cap \mathrm{At}_v(\Pi_1) = f(M) \cap \mathrm{At}_v(\Pi_2)$.*

The proof that $\equiv_m$ is congruent for $\sqcup$ lifts from the case of normal programs (Oikarinen & Janhunen, 2008a) to the disjunctive case using Theorem 5.7.

**Proposition 8.2** *Let $\Pi_1$, $\Pi_2$, and $\Pi$ be DLP-functions. If $\Pi_1 \equiv_m \Pi_2$ and both $\Pi_1 \sqcup \Pi$ and $\Pi_2 \sqcup \Pi$ are defined, then $\Pi_1 \sqcup \Pi \equiv_m \Pi_2 \sqcup \Pi$.*

*Proof.* Let $\Pi_1 = \langle R_1, I_1, O_1, H_1 \rangle$ and $\Pi_2 = \langle R_2, I_2, O_2, H_2 \rangle$ be DLP-functions such that $\Pi_1 \equiv_m \Pi_2$, and $\Pi = \langle R, I, O, H \rangle$ a DLP-function acting as an arbitrary context for $\Pi_1$ and $\Pi_2$ such that $\Pi_1 \sqcup \Pi$ and $\Pi_2 \sqcup \Pi$ are defined. Consider any $M \in \mathrm{SM}(\Pi_1 \sqcup \Pi)$. Theorem 5.7 implies that $M_1 = M \cap \mathrm{At}(\Pi_1) \in \mathrm{SM}(\Pi_1)$ and $N = M \cap \mathrm{At}(\Pi) \in \mathrm{SM}(\Pi)$. Since $\Pi_1 \equiv_m \Pi_2$, we have $I_1 = I_2$, $O_1 = O_2$, and there is a bijection $f : \mathrm{SM}(\Pi_1) \to \mathrm{SM}(\Pi_2)$ such that

$$M_1 \cap (I_1 \cup O_1) = f(M_1) \cap (I_2 \cup O_2) \tag{35}$$

holds for $M_1$. Define $M_2 = f(M_1)$. Since $M_1$ and $N$ are compatible by definition and (35) holds, the models $M_2$ and $N$ are compatible as $I_1 = I_2$ and $O_1 = O_2$. Thus, $M_2 \cup N \in \mathrm{SM}(\Pi_2 \sqcup \Pi)$ by Theorem 5.7 and we have effectively described how $M$ is mapped to a model in $\mathrm{SM}(\Pi_2 \sqcup \Pi)$ by a function $g : \mathrm{SM}(\Pi_1 \sqcup \Pi) \to \mathrm{SM}(\Pi_2 \sqcup \Pi)$ defined by

$$g(M) = f(M \cap \mathrm{At}(\Pi_1)) \cup (M \cap \mathrm{At}(\Pi)).$$

Clearly, $g$ maps the set of visible atoms in $M$ to itself, that is,

$$M \cap (I_1 \cup I \cup O_1 \cup O) = g(M) \cap (I_2 \cup I \cup O_2 \cup O).$$

The justifications for $g$ being a bijection are as follows:

- $g$ is an injection: $M \neq N$ implies $g(M) \neq g(N)$ for all $M, N \in \mathrm{SM}(\Pi_1 \sqcup \Pi)$, since $f(M \cap \mathrm{At}(\Pi_1)) \neq f(N \cap \mathrm{At}(\Pi_1))$ or $M \cap \mathrm{At}(\Pi) \neq N \cap \mathrm{At}(\Pi)$.

- $g$ is a surjection: for any $N \in \mathrm{SM}(\Pi_2 \sqcup \Pi)$, $M = f^{-1}(N \cap \mathrm{At}(\Pi_2)) \cup (N \cap \mathrm{At}(\Pi)) \in \mathrm{SM}(\Pi_1 \sqcup \Pi)$ and $g(M) = N$, since $f$ is a surjection.

The inverse function $g^{-1} : \mathrm{SM}(\Pi_2 \sqcup \Pi) \to \mathrm{SM}(\Pi_1 \sqcup \Pi)$ of $g$ can be defined by setting $g^{-1}(N) = f^{-1}(N \cap \mathrm{At}(\Pi_2)) \cup (N \cap \mathrm{At}(\Pi))$. Thus, $\Pi_1 \sqcup \Pi \equiv_{\mathrm{m}} \Pi_2 \sqcup \Pi$. □

Note that $\Pi \equiv_{\mathrm{m}} \mathrm{GSH}(\Pi)$ follows directly from Theorem 7.6. Applying Proposition 8.2 in the context of Theorem 7.6 indicates that shifting can be localized to a particular component $\Pi_1$ in a larger DLP-function $\Pi_1 \sqcup \Pi$ since $\Pi_1 \sqcup \Pi \equiv_{\mathrm{m}} \mathrm{GSH}(\Pi_1) \sqcup \Pi$.

## 8.2 Verifying Modular Equivalence

Oikarinen and Janhunen (2004) proposed a translation-based method for the verification of *weak equivalence* of disjunctive logic programs. Two logic programs are weakly equivalent iff they have exactly the same set of stable models. Thus, weak equivalence can be seen as a special case of modular equivalence for DLP-functions $\Pi_1$ and $\Pi_2$ where $\mathrm{At}_{\mathrm{i}}(\Pi_1) \cup \mathrm{At}_{\mathrm{h}}(\Pi_1) = \mathrm{At}_{\mathrm{i}}(\Pi_2) \cup \mathrm{At}_{\mathrm{h}}(\Pi_2) = \emptyset$. This motivates us to adjust the translation-based technique for the verification of modular equivalence. As observed in previous work (Janhunen & Oikarinen, 2007; Oikarinen & Janhunen, 2008a), the verification of visible/modular equivalence involves a *counting problem* in general. A reduction of computational time complexity can be achieved for programs that have *enough visible atoms*, referred to as the *EVA property* for short, (Janhunen & Oikarinen, 2007). For any DLP-function $\Pi = \langle R, I, O, H \rangle$, we define the *hidden part* of $\Pi$ as the restricted DLP-function $\Pi_{\mathrm{h}} = \langle \mathrm{Def}_R(H), I \cup O, H, \emptyset \rangle$ which enables the evaluation of hidden atoms in $H$ given arbitrary truth values for all other atoms in $I \cup O$. Recalling Definition 3.2, we use an instantiation of $\Pi_{\mathrm{h}}$ with respect to an interpretation $M_{\mathrm{v}} \subseteq \mathrm{At}_{\mathrm{i}}(\Pi_{\mathrm{h}})$, i.e., $\Pi_{\mathrm{h}}/M_{\mathrm{v}}$, to define the EVA property for the DLP-function $\Pi$.

**Definition 8.3** *A DLP-function* $\Pi = \langle R, I, O, H \rangle$ *has* enough visible atoms *iff* $\Pi_{\mathrm{h}}/M_{\mathrm{v}}$ *has a unique stable model for each* $M_{\mathrm{v}} \subseteq \mathrm{At}_{\mathrm{v}}(\Pi) = \mathrm{At}_{\mathrm{i}}(\Pi_{\mathrm{h}})$.

The idea behind the translation-based method of Oikarinen and Janhunen (2004) is that ordinary disjunctive programs $R_1$ and $R_2$ are weakly equivalent iff their translations $\mathrm{TR}(R_1, R_2)$ and $\mathrm{TR}(R_2, R_1)$ have no stable models. In the following, we propose a modified version of the translation function adjusted to verification of modular equivalence. In order to be able to verify modular equivalence, we need to take the semantics of the atoms in the input signature into account as well as the role of hidden atoms when modular equivalence of programs is under consideration. In the case of DLP-functions, we transform any pair $\Pi_1$ and $\Pi_2$ of *compatible* DLP-functions into a DLP-function $\mathrm{EQT}(\Pi_1, \Pi_2)$ that has a stable model iff there is some stable model $M \in \mathrm{SM}(\Pi_1)$ for which there is no stable model $N \in \mathrm{SM}(\Pi_2)$ with $M \cap \mathrm{At}_{\mathrm{v}}(\Pi_1) = N \cap \mathrm{At}_{\mathrm{v}}(\Pi_2)$. We form the translation as a composition of DLP-functions in order to fully exploit the compositionality of the stable model semantics when justifying the correctness of the method.

In what follows, we use new atoms $a^{\bullet}$, $a^{\circ}$, and $a^{*}$ not appearing in $\mathrm{At}(\Pi_1) \cup \mathrm{At}(\Pi_2)$ for any atom $a$, and we use the shorthand $A^{\bullet} = \{a^{\bullet} \mid a \in A\}$ for any set $A$ of atoms, and analogously defined shorthands $A^\circ$ and $A^*$. Moreover, diff, unsat, unsat$^\bullet$, and ok are new atoms not appearing in $\mathrm{At}(\Pi_1) \cup \mathrm{At}(\Pi_2)$. The translation $\mathrm{EQT}(\Pi_1, \Pi_2)$, which is to be summarized by Definition 8.4 below, consists of the following three parts:

(i) The DLP-function $\Pi_1$ naturally captures a stable model $M \in \mathrm{SM}(\Pi_1)$.

(ii) The DLP-function $\mathrm{hidden}(\Pi_2) = \langle R_\mathrm{h}, I \cup O, H^*, \emptyset \rangle$ provides a representation for the hidden part of $\Pi_2 = \langle R, I, O, H \rangle$ evaluated with respect to the visible part of $M$. The input signature of $\mathrm{hidden}(\Pi_2)$ consists of the visible atoms in $\mathrm{At_v}(\Pi_2) = \mathrm{At_v}(\Pi_1) = I \cup O$. The set $R_\mathrm{h}$ contains a rule $A_\mathrm{h}^* \leftarrow B_\mathrm{v} \cup B_\mathrm{h}^*, \sim(A_\mathrm{v} \cup C_\mathrm{v} \cup C_\mathrm{h}^*)$ for each $A \leftarrow B, \sim C \in R$ such that $A_\mathrm{h} \neq \emptyset$, i.e., $A \leftarrow B, \sim C \in \mathrm{Def}_R(H)$. The hidden parts of rules are renamed systematically using atoms from $\mathrm{At_h}(\Pi_2)^*$. This is to capture the unique stable model $N$ for $(\Pi_2)_\mathrm{h}/M_\mathrm{v}$ expressed in $\mathrm{At_h}(\Pi_2)^*$ rather than $\mathrm{At_h}(\Pi_2)$. Note that the *existence and uniqueness* of such an $N$ is guaranteed by the EVA property.

(iii) Finally, the DLP-function

$$\mathrm{TR}(\Pi_2) = \langle R_\mathrm{TR}, I \cup O \cup H^*, O^\bullet \cup H^\bullet \cup \{\mathsf{unsat}, \mathsf{unsat}^\bullet, \mathsf{diff}, \mathsf{ok}\}, O^\circ \cup H^\circ \rangle$$

provides a minimality check. The set $R_\mathrm{TR}$ contains

1. a rule $\mathsf{unsat} \leftarrow B_\mathrm{v} \cup B_\mathrm{h}^*, \sim(A_\mathrm{v} \cup A_\mathrm{h}^* \cup C_\mathrm{v} \cup C_\mathrm{h}^*)$ for each rule $A \leftarrow B, \sim C \in R$,

2. rules $a^\bullet \leftarrow a, \sim a^\circ, \sim\mathsf{unsat}$ and $a^\circ \leftarrow a, \sim a^\bullet, \sim\mathsf{unsat}$ for each $a \in O$, and rules $a^\bullet \leftarrow a^*, \sim a^\circ, \sim\mathsf{unsat}$ and $a^\circ \leftarrow a^*, \sim a^\bullet, \sim\mathsf{unsat}$ for each $a \in H$,

3. a rule $\mathsf{unsat}^\bullet \leftarrow B_\mathrm{i} \cup B_\mathrm{o}^\bullet \cup B_\mathrm{h}^\bullet, \sim(A_\mathrm{i} \cup A_\mathrm{o}^\bullet \cup A_\mathrm{h}^\bullet \cup C_\mathrm{v} \cup C_\mathrm{h}^*), \sim\mathsf{unsat}$ for each rule $A \leftarrow B, \sim C \in R$,

4. a rule $\mathsf{diff} \leftarrow a, \sim a^\bullet, \sim\mathsf{unsat}$ for each $a \in O$, and a rule $\mathsf{diff} \leftarrow a^*, \sim a^\bullet, \sim\mathsf{unsat}$ for each $a \in H$, and

5. the following rules:

$$\mathsf{ok} \leftarrow \mathsf{unsat}, \quad \mathsf{ok} \leftarrow \mathsf{diff}, \sim\mathsf{unsat}, \sim\mathsf{unsat}^\bullet, \quad \text{and} \quad \bot \leftarrow \sim\mathsf{ok}.$$

The intuition behind the translation $\mathrm{TR}(\Pi_2)$ is as follows. The rules in the first item check whether an interpretation $L \subseteq \mathrm{At}(\Pi_2)$ corresponding to the actual input $K = (L \cap (I \cup O)) \cup \{a^* \mid a \in L \cap H\} \subseteq \mathrm{At_v}(\Pi_2) \cup \mathrm{At_h}(\Pi_2)^*$ for $\mathrm{TR}(\Pi_2)$ satisfies the rules in $\Pi_2$. If the rules of $\Pi_2$ are satisfied, then the rules in items 2–4 are activated by the literals $\sim\mathsf{unsat}$ in their bodies. The rules in the second item are used to generate a subset $L'$ of $L$ such that $L' \cap \mathrm{At_i}(\Pi_2) = L \cap \mathrm{At_i}(\Pi_2)$. This is achieved by introducing a new atom $a^\bullet$ for each $a \in \mathrm{At_o}(\Pi_2) \cup \mathrm{At_h}(\Pi_2)$. The rules in the third item check whether the representation of $L'$ in $\mathrm{At_i}(\Pi_2) \cup \mathrm{At_o}(\Pi_2)^\bullet \cup \mathrm{At_h}(\Pi_2)^\bullet$, i.e., $K' = (L' \cap I) \cup \{a^\bullet \mid a \in L' \cap (O \cup H)\}$, satisfies the rules in $\Pi_2^L$. The rules in the fourth item check whether $L'$ is a proper subset of $L$. Finally, the rules in the fifth item summarize the reasons why $L$ cannot be a stable model of $\Pi_2$, i.e., either the rules in $\Pi_2$ are not satisfied in $L$, or $L$ is not a minimal model of $\Pi_2^L$. As the net effect of this construction, $\mathrm{TR}(\Pi_2)/K$ has a stable model iff $L$ is not a stable model of $\Pi_2$.

**Definition 8.4** *Let $\Pi_1$ and $\Pi_2 = \langle R, I, O, H \rangle$ be compatible DLP-functions having enough visible atoms. Then, the translation $\mathrm{EQT}(\Pi_1, \Pi_2)$ is given by $\Pi_1 \sqcup \mathrm{hidden}(\Pi_2) \sqcup \mathrm{TR}(\Pi_2)$.*

The translation $\mathrm{TR}(\Pi_2)$ for the minimality check essentially contains the same rules as $\mathrm{TR}(R_1, R_2) \setminus R_1$, where $\mathrm{TR}(R_1, R_2)$ is the translation defined by Oikarinen and Janhunen (2004) for sets $R_1$ and $R_2$ of disjunctive rules. There are two further aspects, however. First, occurrences of hidden atoms from $H$ are additionally represented using their counterparts from $H^*$. Second, we only need renamed versions of atoms in $O \cup H$ because the interpretation of atoms in the input signature $I$ is kept fixed. Finally, we note that for DLP-functions $\Pi_1$ and $\Pi_2$ which correspond to ordinary disjunctive logic programs, i.e., for $\Pi_1 = \langle R_1, \emptyset, O, \emptyset \rangle$ and $\Pi_2 = \langle R_2, \emptyset, O, \emptyset \rangle$, the translation $\mathrm{EQT}(\Pi_1, \Pi_2)$ coincides with $\mathrm{TR}(R_1, R_2)$.

**Theorem 8.5** *Let $\Pi_1$ and $\Pi_2$ be compatible DLP-functions having enough visible atoms. Then, $\Pi_1 \equiv_\mathrm{m} \Pi_2$ iff both $\mathrm{SM}(\mathrm{EQT}(\Pi_1, \Pi_2)) = \emptyset$ and $\mathrm{SM}(\mathrm{EQT}(\Pi_2, \Pi_1)) = \emptyset$.*

*Proof sketch.* Let $\Pi_1$ and $\Pi_2 = \langle R, I, O, H \rangle$ be compatible DLP-functions having enough visible atoms. By Theorem 5.7, given compatible interpretations $M_1 \subseteq \mathrm{At}(\Pi_1)$, $M_2 \subseteq \mathrm{At}(\mathrm{hidden}(\Pi_2))$, and $M_3 \subseteq \mathrm{At}(\mathrm{TR}(\Pi_2))$, $M = M_1 \cup M_2 \cup M_3$ is a stable model of the translation $\mathrm{EQT}(\Pi_1, \Pi_2)$ iff $M_1 \in \mathrm{SM}(\Pi_1)$, $M_2 \in \mathrm{SM}(\mathrm{hidden}(\Pi_2))$, and $M_3 \in \mathrm{SM}(\mathrm{TR}(\Pi_2))$. Given any interpretation $M_1 \subseteq \mathrm{At}(\Pi_1)$, there is a unique stable model $M_2 \in \mathrm{SM}(\mathrm{hidden}(\Pi_2))$ compatible with $M_1$, since $\Pi_2$ has the EVA property. Hence, $\mathrm{hidden}(\Pi_2)$ does not constrain stable models in the composition $\mathrm{EQT}(\Pi_1, \Pi_2)$. Whenever $M_3$ is compatible with both $M_1$ and $M_2$, it holds that $M_3 \cap (I \cup O \cup H^*) = (M_1 \cup M_2) \cap (I \cup O \cup H^*)$ and $M_3 \in \mathrm{SM}(\mathrm{TR}(\Pi_2))$ iff the interpretation $M_3 \cap (I \cup O) \cup \{a \in H \mid a^* \in M_3\}$ is not a stable model of $\Pi_2$ as established by Oikarinen and Janhunen (2004, Theorem 1). □

When verifying modular equivalence of DLP-functions of the forms $\Pi_1 \sqcup \Pi$ and $\Pi_2 \sqcup \Pi$, it is possible to further streamline the translations involved in the verification task.

**Theorem 8.6** *Let $\Pi_1$ and $\Pi_2$ be compatible DLP-functions having enough visible atoms, and $\Pi$ a DLP-function such that $\Pi_1 \sqcup \Pi$ and $\Pi_2 \sqcup \Pi$ are defined. Then, $\Pi_1 \sqcup \Pi \equiv_\mathrm{m} \Pi_2 \sqcup \Pi$ iff both $\mathrm{SM}(\mathrm{EQT}(\Pi_1, \Pi_2) \sqcup \Pi) = \emptyset$ and $\mathrm{SM}(\mathrm{EQT}(\Pi_2, \Pi_1) \sqcup \Pi) = \emptyset$.*

The context $\Pi$ can be an arbitrary DLP-function, i.e., it is not necessary for $\Pi$ to have the EVA property, as long as $\Pi_1 \sqcup \Pi$ and $\Pi_2 \sqcup \Pi$ are defined. To prove Theorem 8.6, notice that due to the structure of the translation, $\mathrm{EQT}(\Pi_1, \Pi_2) \sqcup \Pi$ is defined whenever $\Pi_1 \sqcup \Pi$ is defined, and then Theorems 5.7 and 8.5 can be applied.

## 9. Related Work

Eiter et al. (1997a) consider the use of *disjunctive datalog programs* as *query programs* over relational databases. In their approach, query programs are formalized as triples $\langle \pi, \mathbf{R}, \mathbf{S} \rangle$ where $\pi$ is a set of disjunctive rules and $\mathbf{R}$ and $\mathbf{S}$ are the signatures for the input and output relations, respectively, whereas auxiliary (hidden) predicates are left implicit. Hence, in the propositional case, the only notable difference with respect to Definition 2.1 is that input atoms are not allowed to occur in the heads of disjunctive rules. As regards semantics, the program $\pi$ is reduced with respect to a complete input database $D$ specified in terms of **R**, yielding the instantiation $\pi[D]$, and, among others, stable-model semantics is applied to $\pi[D]$ in analogy to Definition 3.2. However, in contrast to our modular architecture, Eiter et al. (1997a) take both *positive and negative dependencies* into account and no recursion between modules is tolerated. The resulting hierarchy of *complete components* admits a straightforward generalization of the *splitting sequences* (Lifschitz & Turner, 1994). The essential difference is that a partial order rather than a total order of modules is assumed. In this respect, it is worth pointing out that partial orders of DLP-functions are permitted by $\sqcup$.

Modularity has gained more attention in the context of conventional (monotonic) logic programming; see the work of Bugliesi, Lamma, and Mello (1994) for a survey. Two mainstream approaches are identified: The first is called *programming-in-the-large* in which algebraic operators are introduced for the construction of logic programs out of modules. The approach of our paper falls into this category—the join $\sqcup$ being an example of such operators. The other, and quite different *programming-in-the-small* approach, is to extend the underlying logical language in terms of abstraction mechanisms. In the approach of Eiter et al. (1997b), for instance, logic program modules are viewed as *generalized quantifiers* which are allowed to nest but only in a hierarchical fashion. To give an idea of this approach, consider a module that formalizes the transitive closure of a relation denoted by a predicate $\mathsf{rel}(\cdot, \cdot)$:

$$\mathsf{tclo}(x,y) \leftarrow \mathsf{rel}(x,y); \quad \mathsf{tclo}(x,y) \leftarrow \mathsf{tclo}(x,z), \mathsf{rel}(z,y).$$

Here, $\mathsf{tclo}(\cdot, \cdot)$ acts as the *output predicate* of the module $\mathsf{tclo}[\mathsf{rel}]$ whereas $\mathsf{rel}(\cdot, \cdot)$ is its only *input predicate*. The module can be invoked to create the transitive closure of any binary relation substituted for $\mathsf{rel}(\cdot, \cdot)$ above. Consider, for instance, the rule

$$\mathsf{loop}(x) \leftarrow \mathsf{tclo}[\mathsf{edge}](x,y), \mathsf{tclo}[\mathsf{edge}](y,x)$$

which captures nodes involved in the loops of a directed graph whose edges are supposed to be represented with the predicate $\mathsf{edge}(\cdot, \cdot)$. In our approach, the call $\mathsf{tclo}[\mathsf{edge}]$ would result in one module as part of the respective ground program with input and output signatures

$$I_n = \{\mathsf{edge}(x,y) \mid 1 \leq x,y \leq n\} \text{ and } O_n = \{\mathsf{tclo}(x,y) \mid 1 \leq x,y \leq n\}$$

in the case of $n$ vertices. However, in the architecture of Eiter et al. (1997b), the module $\mathsf{tclo}[\mathsf{rel}]$ can be invoked several times to form transitive closures of different relations. In our effectively propositional approach, each invocation of $\mathsf{tclo}[\mathsf{rel}]$ would map to a new module. Although these modules could be obtained by straightforward renaming of predicates, this aspect illustrates the power of the programming-in-the-small approach. Here, $\mathsf{tclo}[\mathsf{rel}]$ acts as a new parameterized connective which the programmer can concisely refer to as a new relation, viz. the transitive closure of $\mathsf{rel}$ in this case. But, in spite of succinctness at this point, such relations may have to be unwound in an actual implementation. This aspect is made explicit in the modular action description (MAD) language proposed by Lifschitz and Ren (2006): a modular action description is turned into a single-module description in a recursive fashion. The outcome determines the meaning of the modular description via an embedding into ASP (Lifschitz & Turner, 1999).

Faber, Greco, and Leone (2007) apply the *magic-set method* in the evaluation of datalog programs with negation. Their notion of a module is based on the concept of an *independent* *set.* For a non-disjunctive logic program $\Pi = \langle R, \emptyset, O, \emptyset \rangle$, such a set $S \subseteq O$ satisfies, for any $a \in S$, the following two conditions:

1. if there is a rule $h \leftarrow B, \sim C \in R$ such that $h = a$, then $B \cup C \subseteq S$, and

2. if $a \in B \cup C$ for some *dangerous rule* $h \leftarrow B, \sim C \in R$, then $\{h\} \cup B \cup C \subseteq S$.

We skip the exact definition of dangerous rules which, roughly speaking, may interfere with the existence of stable models. It is clear that independent sets are splitting sets in the sense of Definition 5.15, but not vice versa in general. Hence, the module theorem provided by Faber et al. (2007) can be viewed as a special case of the splitting-set theorem and, therefore, observations presented in Section 5.3 apply to independent sets as well.

## 10. Conclusion and Discussion

In this paper, we introduced a formal framework for modular programming in the context of disjunctive logic programs under stable-model semantics. The framework is based on the notion of a *DLP-function* which puts into effect appropriate input/output interfacing for disjunctive logic programs. Analogous module concepts have already been studied in the cases of normal logic programs and SMODELS programs (Oikarinen & Janhunen, 2008a) and even propositional theories (Janhunen, 2008a), but the special characteristics of disjunctive rules are properly taken into account in the syntactic and semantic definitions of DLP-functions presented herein. In this respect, we would like to draw the reader's attention to Definition 2.1 (item 2), Definition 2.2 (items 4–5), as well as Definition 3.2.

Undoubtedly, the main result of this paper is the module theorem, i.e., Theorem 5.7, which is proved for DLP-functions in general—thus covering the class of disjunctive programs. The module theorem is important as it provides a compositional semantics for disjunctive programs and it generalizes existing approaches such as those based on *splitting sets* (Lifschitz & Turner, 1994) and *magic sets* (Faber et al., 2007). Although our approach is based on a number of design decisions, e.g., as regards the definition of module composition, it nevertheless brings out the limits of modular programming in the context of a nonmonotonic declarative language. The module theorem can be exploited in a number of ways in ASP based on disjunctive logic programs. As demonstrated in Section 6, it provides the basis for decomposing disjunctive programs into their components and hence the localization of reasoning tasks. Moreover, as established in Section 7, the technique of shifting disjunctive rules can be generalized for disjunctive programs involving head-cycles. Actually, the generalized form enables us to remove shared disjunctive rules altogether but this might not be desirable due to higher space requirements. Finally, the theory of modular equivalence is fully applicable to DLP-functions as demonstrated in Section 8.

In addition to the results discussed above, we anticipate further applications of the module theorem in the future. We strongly believe that research in this direction not only yields results of theoretical interest but also leads to the development of practically useful software engineering methods for ASP. In fact, first tools for decomposing and linking programs have already been implemented in the context of the SMODELS system.[14] The results of Section 6 enable the development of analogous tools to be used with disjunctive solvers such as CLASPD, CMODELS, DLV, and GNT. There is also an implementation of the general shifting principle, called DENCODE, in the ASP tool collection. The results of Section 8 pave the way for extending a translation-based verification tool, DLPEQ (Janhunen & Oikarinen, 2004), for the verification of modular equivalence. Such an extension is already available in the respective tool, LPEQ, for SMODELS programs (Oikarinen & Janhunen, 2009).[15]

---

14. See MODLIST and LPCAT in the ASP tools collection at `http://www.tcs.hut.fi/Software/asptools/`.

**Acknowledgments** This work was partially supported by the Academy of Finland under projects #211025 ("Advanced Constraint Programming Techniques for Large Structured Problems") and #122399 ("Methods for Constructing and Solving Large Constraint Models"), and by the Austrian Science Foundation (FWF) under projects P18019 ("Formal Methods for Comparing and Optimizing Nonmonotonic Logic Programs") and P21698 ("Methods and Methodologies for Developing Answer-Set Programs"). The authors would like to thank the anonymous referees for their constructive comments as well as Martin Gebser and Torsten Schaub for their suggestion to exploit program completion and loop formulas in the proof of the module theorem. A preliminary version of this paper appeared in the proceedings of the 9th International Conference on Logic Programming and Nonmonotonic Reasoning (LPNMR'07), Vol. 4483 of LNCS, pp. 175–187, Tempe, AZ, USA, Springer.

---

15. Verification tools mentioned here are available at `http://www.tcs.hut.fi/Software/lpeq/`.